\documentclass[prx,twocolumn,floatfix,nobibnotes,superscriptaddress]{revtex4-1}

\usepackage[fleqn]{amsmath}
\usepackage[tight]{subfigure}
\usepackage[english]{babel}
\usepackage{csquotes}
\usepackage{amssymb}
\usepackage{amsmath}
\usepackage{braket}
\usepackage{braket}
\usepackage{graphicx,color}
\usepackage{epstopdf}
\usepackage{dcolumn}
\usepackage{mathtools}
\usepackage{amsmath}
\usepackage{physics}
\usepackage{accents}
\usepackage{mathrsfs}
\usepackage{fancyref}
\usepackage{siunitx}
\usepackage{float}

\usepackage[citecolor=blue,colorlinks=true,linkcolor=blue]{hyperref}

\begin{document}

\title{A multi-scale approach for magnetisation dynamics: Unraveling exotic magnetic states of matter}

\author{\'{E}. M\'{e}ndez}
\affiliation{Division of Materials theory, Department of Physics and Astronomy, Uppsala University, Box 516, 75121 Uppsala, Sweden}
\author{M. Poluektov}
\affiliation{International Institute for Nanocomposites Manufacturing, WMG, University of Warwick, Coventry CV4 7AL, UK}
\author{G. Kreiss}
\affiliation{Department of Information Technology, Uppsala University, Box 337, SE-751 05 Uppsala, Sweden}
\author{O. Eriksson}
\affiliation{Division of Materials theory, Department of Physics and Astronomy, Uppsala University, Box 516, 75121 Uppsala, Sweden}
\affiliation{School of Science and Technology, \"Orebro University, SE-70182 \"Orebro, Sweden}
\author{M. Pereiro}
\affiliation{Division of Materials theory, Department of Physics and Astronomy, Uppsala University, Box 516, 75121 Uppsala, Sweden}

\begin{abstract}
Crystallographic lattice defects strongly influence dynamical properties of magnetic materials at both microscopic and macroscopic length scales. A multi-scale approach to magnetisation dynamics, which is presented in this paper, accurately captures such effects. The method is illustrated using examples of systems with localized, non-trivial topological properties, e.g. in the form of skyrmions and chiral domain walls that interact with lattice dislocations. Technical aspects of the methodology involve multi-scale magnetisation dynamics that connects atomistic and continuum descriptions. The technique is capable of solving the Landau-Lifshitz-Gilbert equations efficiently in two regions of a magnetic material --- the mesoscopic and the atomistic regions, which are coupled in a seamless way. It is demonstrated that this methodology allows simulating realistically-sized magnetic skyrmions interacting with material defects and novel physical effects, uncovered using this theoretical methodology, are described.
\end{abstract}

\maketitle

\section*{Introduction}
\label{sec:intro}

It is of paramount importance in material science to simulate the behaviour of matter both at small and large length scales. New and exciting macroscopic phenomena in materials are usually driven by the perturbation of physical properties at local and very small length scales. For example, dislocations or defects of atomic positions influence plastic deformation of materials and in magnetic media, they are a source of domain wall pinning and the Barkhausen effect. In many cases, local properties cannot be disentangled from the global properties of the material, in which they are embedded, and, consequently, both spatial scales are required to be treated on equal footing. This is a highly non-trivial task, which this article addresses. Technological applications, e.g. the promising area of spintronics \cite{vzutic2004spintronics} and the evolving field of quantum computing \cite{burkard2000spintronics}, bring relevance to the field of magnetism. Scientific advancements in this area are provided by several new experimental techniques, e.g. pump-probe measurements of magnetism\cite{beaurepaire}. For interpretation of experiments, simulations are highly relevant and allows understanding or even predicting new and interesting phenomena. In essentially all fields of physics, simulation tools are used in parallel to experimental observations. 

Magnetic phenomena can be studied at a variety of scales, from macroscopic to atomic. Different models are traditionally used at each level in order to consider specific physical phenomena that are significant for a given time and length scale. Two commonly used paradigms are the micromagnetic model and the atomic description that is based on the Heisenberg Hamiltonian. Micromagnetic theory \cite{abert} describes phenomena typically at micrometre length scale. This methodology is well established and has been used for several decades. It is known to provide results of a relatively high degree of agreement with experimental data, at least for the length scale and time range it is designed for. Fundamental to this model is the assumption that the magnetisation varies smoothly in space and micro-structural information is embedded into effective parameters. These conditions impose a limit on the applicability of this model to small magnetic structures or intricate atomistic arrangements. On the other hand, the Heisenberg spin model provides a discrete, atomic-scale description of magnetism. This model is capable of representing magnetisation dynamics on an atomic length scale, naturally including effects of lattice irregularities \cite{SDbook}. However, the computational cost of atomistic spin dynamics simulations makes it unfeasible for large computational domains. Therefore, it has difficulties describing the properties of samples at sizes that are used in some experimental measurements.

Micromagnetic models allow distributing computational points in a controlled manner. These methods have been reported in the literature and are based on finite differences, e.g. in Ref.~\cite{garcia2006adaptive,kim2017mimetic}, or finite elements, e.g. in Ref.~\cite{tako1997finite}. The atomistic model does not permit redistributing computational points, but it can be used in combination with micromagnetic models leading to various multi-scale schemes. There are several versions of such coupling schemes developed in application to magnetic materials \cite{GarciaSanchez2005,Jourdan2008,andreas2014multiscale,de2016multiscale} and the reader is referred to a review article published in Ref.~\cite{Hertel2018} for an exhaustive overview. 

The aim of this paper is to demonstrate the applicability of a multi-scale method to analysing complex magnetisation dynamics in systems where large and small length scales play role, e.g. skyrmion dynamics and its influence from defects on the atomic scale. Limitations of this methodology are established by considering a set of examples that are challenging for any similar numerical method that aims at simulating magnetisation dynamics of materials. To perform the investigation, an implementation of a multi-scale method has been made in the UppASD software \cite{skubic2008method,webUppasd}, taking on-board techniques from Refs.~\cite{Poluektov2018,Poluektov2016}. The micromagnetic model is shown here to have sufficient accuracy for regions where the magnetisation varies smoothly, while atomic-level simulations can reproduce the much more complex behaviour at shorter length scales. Framing a small atomistic region with a thick micromagnetic layer in a way proposed in this paper, permits estimating how perturbations propagate in the material, while avoiding reflections that pollute the region of interest and keeping the computational effort and accuracy mainly within the atomistic part. As is demonstrated below, the multi-scale method outlined here is able to simulate dynamics of topological excitations, e.g. skyrmions with realistic (experimental) size, and their interaction with atomistic defects, such as dislocations.

\section*{Methods}

The type of atomistic-continuum coupling considered in this paper is the partitioned-domain method. In this scheme, the computational domain is split into regions that belong to different models, with an explicit interface between the regions. There are two conceptually distinct atomistic-continuum coupling methodologies --- the energy-based and the force-based coupling \cite{Tadmor2011}. There are two major challenges for partitioned-domain atomistic-continuum coupling methods: handling non-local atomistic interactions, which is relevant for the energy-based methods applied in both statics and dynamics, and averaging high-frequency excitations within the atomistic solution that must not reflect from the boundary between the atomistic and the coarsely-discretised continuum regions. It must be emphasised that the latter is not an issue of coupling methods and results from disparate discretisations of the regions. Atomistic-continuum coupling methods originate from the field of mechanics and date back to 1990s and, therefore, the highlighted issues were largely addressed in application to modelling deformation, e.g. Ref.~\cite{Ortner2014} solves the local/non-local coupling issue in the energy-based methods in mechanics and Ref.~\cite{Qu2005} proposes a way of dealing with high-frequency atomistic excitations. In the field of magnetism, the local/non-local coupling has been addressed in Ref.~\cite{Poluektov2018}, while a technique for averaging high-frequency spin motion at the interface has been proposed in Ref.~\cite{Poluektov2016}. 

The technical aspects of the method proposed and used here involves the merging of an atomistic and a continuum description of the magnetisation dynamics into an established software --- UppASD \cite{webUppasd}. Most of the technical aspects used in this work have been introduced previously \cite{skubic2008method,SDbook,Poluektov2016,Poluektov2018} and, therefore, relevant technical aspects of this work are presented in Supplementary Note S1.

\section*{Results}

It is well-known that it is not possible to construct a completely error-free atomistic-continuum coupling, as discussed in Ref.\cite{Tadmor2011}. Previous studies \cite{Poluektov2018,Poluektov2016} focused on analysing the errors introduced by the coupling approach both analytically and computationally. The aim of the examples of this paper is to show that it is possible to construct the coupling such that atomistic and continuum regions behave in a similar way and the errors introduced by the atomistic-continuum interface are sufficiently small and do not influence the microscopic and the macroscopic behaviours. In the general case, the atomistic-continuum coupling errors depend on the continuum discretisation and on the relation of magnetisation gradient to atomistic lattice spacing \cite{Poluektov2018}. 

In the computational examples and figures, Cartesian coordinates $x$, $y$, $z$ correspond to the horizontal, the vertical and the out-of-plane axes, respectively. In figures, the red and the blue colours indicate magnetisation pointing in-plane and out-of-plane, respectively.

\subsection{Domain wall motion}

\begin{figure}[tb]
  \begin{center}
    \includegraphics[width=0.475\textwidth]{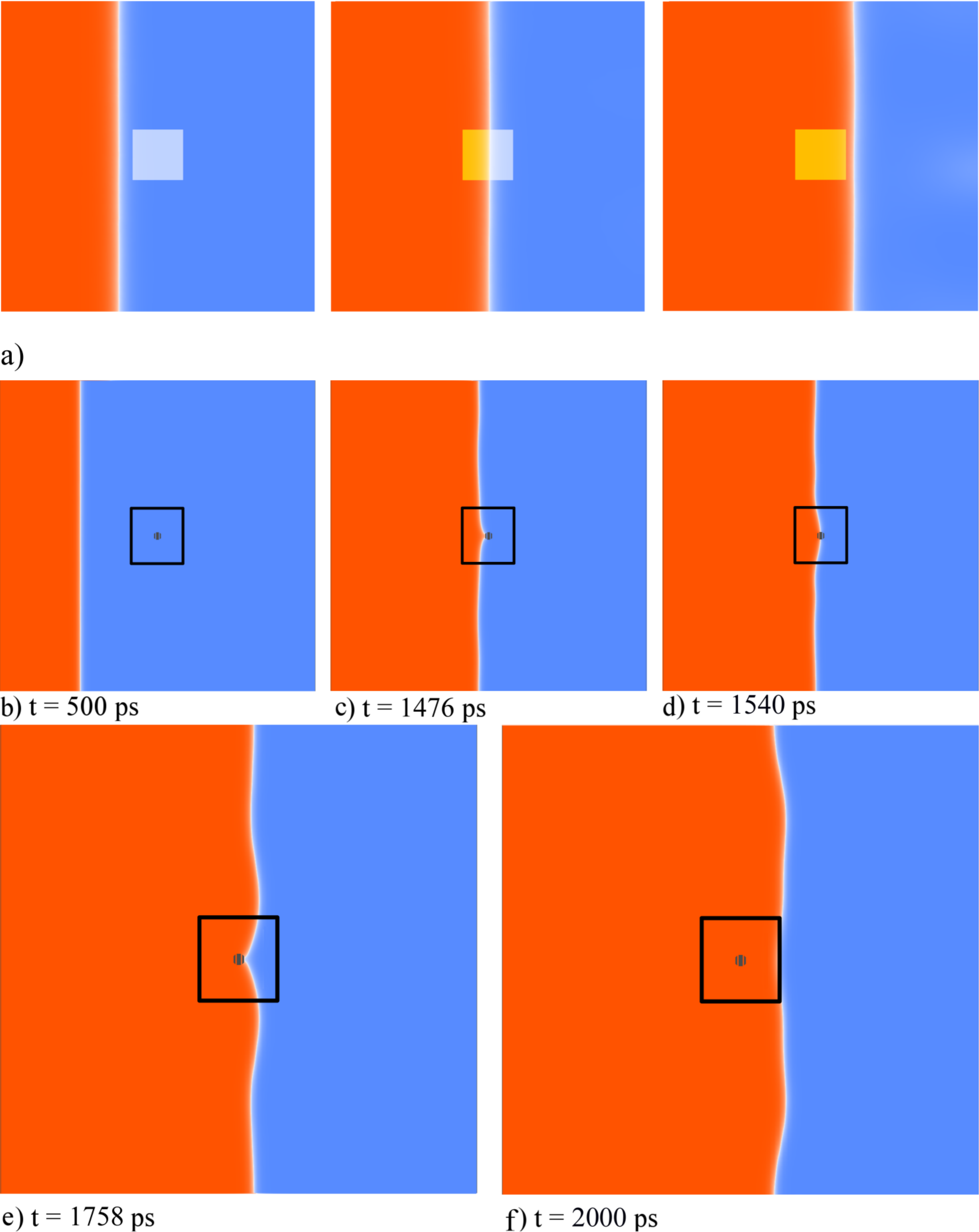}
  \end{center}
  \caption{a) Domain wall crossing atomistic region, which is indicated by the white/yellow square. b)-f) Snapshots at successive steps of the time evolution of the domain wall crossing an atomistic region containing a round hole, which is indicated by the black colour. The atomistic region is indicated by the black frame.}
  \label{fig1}
\end{figure}

A basic domain wall motion, with a Hamiltonian that involves Heisenberg exchange and uniaxial anisotropy, is considered. The simulations consider two regions of opposing magnetisation, each having orientation parallel to the lowest energy direction, specified by the easy axis \cite{note}. The precise material parameters for the simulations of domain wall motion are detailed in Supplementary Note S3a). In this example, the domain wall width is approximately $50$ interatomic distances. The simulation shown in Fig.~\ref{fig1}a) shows that the domain wall, which starts in the continuum region, is not modified or distorted when it reaches the atomistic region --- white/yellow box in the centre of Fig.~\ref{fig1}a). This illustrates that the handshake between the two regions is seamless and that an object, which is not expected to change its shape, correctly maintains its original configuration when moving across interfaces between the regions. The same conclusion is reached when considering the domain wall that is leaving the atomistic region and entering the continuum region. The dynamics of the domain wall is found in Supplementary Video 1.

\subsection{Domain wall motion and the Barkhausen effect}

The cause of the Barkhausen effect has been discussed previously in Ref.~\cite{aharoni}. Materials usually contain irregularities in the crystallographic lattice, such as dislocations, stacking faults, impurity atoms, vacancies and lattice distortions due to micro-stresses. A domain wall typically needs a local increase in energy around the defect to cross one of these irregularities and, therefore, becomes pinned --- parts of the domain wall near the defect lag behind the rest of the domain wall. The unpinning of the domain wall causes a detectable change in the magnetic flux, which was picked up in Barkhausen's coils. 

The numerical examples investigated here consider defects in the form of microscopically small, regular shaped holes in the hexagonal crystal lattice. Snapshots for different times of the domain wall motion are shown in Fig.~\ref{fig1}b)-f) and are seen to represent the wall moving over the defect. Regions of the domain wall that come close to the hole move quickly towards the hole and get stuck around it, as can be seen in Fig.~\ref{fig1}c). The domain wall is stuck for an appreciable time, even if the wall in the rest of the material keeps moving, as can be seen in Fig.~\ref{fig1}e). Eventually, in just a relatively short period of time ($\approx\! 200\! -\! 500\,\mathrm{ps}$) a substantial amount of the material changes its magnetisation. In an experiment, such fast changes of the magnetisation would lead to big changes in the magnetic flux that is connected to the Barkhausen effect (see Supplementary Video 2). Although micromagnetic models have successfully discussed the Barkhausen effect \cite{chubykalo} and generate results similar to those of Fig.~\ref{fig1}, the approach adopted here has greater descriptive possibilities due to treatment of the defect region in an atomistic way, with unique properties of each atom around the defect and with parameters of the spin Hamiltonian that are evaluated from the first principles theory.

\subsection{Topological magnetic states}

In condensed matter physics, magnetic skyrmions are spatially localized excitations that preserve their shape. These skyrmions are homotopically distinct from the ferromagnetic state, meaning that there exists no continuous transformation that transforms the magnetic texture of a skyrmion to the ferromagnetic state. This topological argument has been discussed critically in the literature, particularly with respect to stability and the possibility of using these objects as carriers of information \cite{pereiro,pereiro1}. Skyrmions have been observed experimentally, both in static and dynamic conditions \cite{hoffmann}. 

Magnetic skyrmions come in two versions --- hedgehog skyrmions have their moments pointing in the radial direction, while vortex skyrmions have their moments circulating around the centre. A particular type of skyrmion depends on the direction of $\mathbf{D}_{ij}$ vector with respect to vector $\mathbf{r}_{ij}$ connecting atomic positions $i$ and $j$. The material parameters used in the subsequent simulations, with the exception of skyrmion-dislocation interactions, are listed in Supplementary Note S3b). These parameters correspond to a one-atom thick layer of iron resting atop a substrate of iridium. It has already been demonstrated experimentally that skyrmions exit in this system \cite{heinze2011spontaneous}.

\subsubsection{Spin spiral and Skyrmion lattice}

Considering only the Heisenberg exchange, energy is lowered for an atom and its neighbours when their atomic moments become parallel. In contrast, the energy associated to the Dzyaloshinskii-Moriya interaction can be minimized when the magnetic moments of neighbouring atoms are non-collinear and the minimum energy is found when moments form an angle of $90$ degrees. In materials where both terms contribute significantly, a competition between interactions may introduce many different low-energy states. One family of states of these magnets is known as the spin spiral. For these systems, locally, the variation of spins becomes relatively small in one direction and relatively large in an orthogonal direction. There is a tendency for directions of minimal variation to align, creating long lines of aligned magnetisation. In this example, the multi-scale approach is used to describe this class of materials. 

The time evolution of the magnetic moments achieving the spin spiral state is shown in Fig.~\ref{fig2}a)-c) and in Supplementary Video 3. Initially, the system is thermalized by using an annealing process --- the temperature is decreased gradually starting from \SI{100}{K} until it reaches \SI{0.1}{K}. After the thermalisation process, the system obtained a ferromagnetic state with all moments pointing along the $z$-direction (data not shown). Afterwards, the magnetic moments are allowed to evolve in time at $T=0.1\,\mathrm{K}$, until the system converges into a spin spiral state, as shown in Fig.~\ref{fig2}a)-c). The stripes of opposing magnetisation are easy to identify by the $z$-component of the moments. If an external field of a sufficient magnitude is applied along the $z$-axis to the spin-spiral state (\SI{5}{T} in the example here), skyrmions may appear in the material, forming what is known as a skyrmion lattice, see Supplementary Video 4 and Fig.~\ref{fig2}d)-f). In simulations of both the skyrmion lattice and the spin spiral configuration, the border between the micromagnetic and atomistic region is completely transparent. This implies that the interface between two regions does not introduce any spurious errors and the time evolution of the magnetisation behaves according to expectations in both regions --- the size and the shape of skyrmions is the same in both regions.

\begin{figure}[tb]
  \begin{center}
    \includegraphics[width=0.48\textwidth]{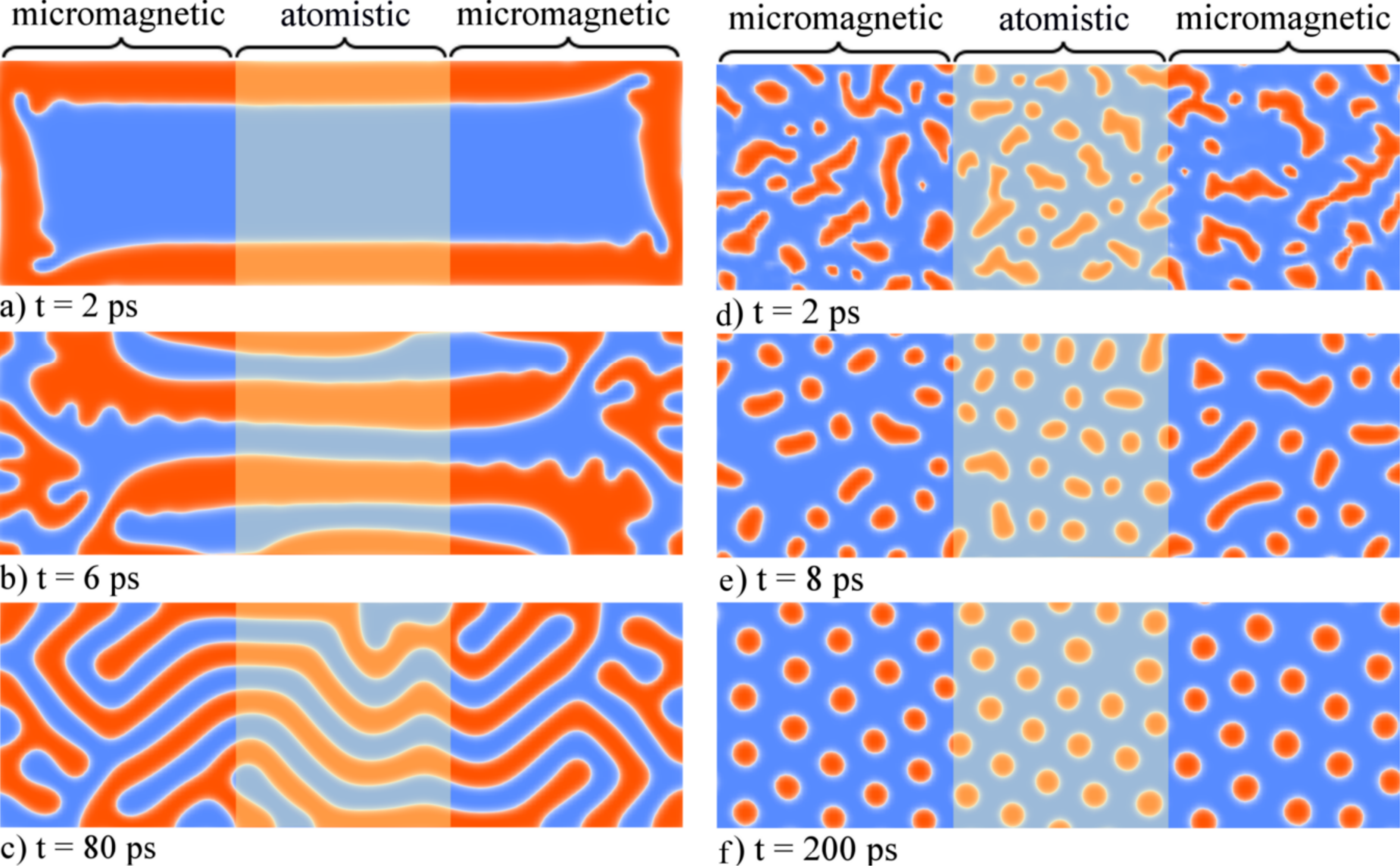}
  \end{center}
  \caption{a)-c) Three snapshots of the time evolution of the magnetisation of one atom thick layer of Fe on Ir (111). The ferromagnetic initial state (not shown) evolves in time to a spin spiral configuration. d)-f) The application of an external magnetic field perpendicular to the plane of the sample drives the spin-spiral state shown in c) to a Skyrmion lattice state. Blue and red regions correspond to the magnetisation pointing mainly along and opposite the $z$-direction, respectively.}
  \label{fig2}
\end{figure}

\begin{figure}[tb]
  \begin{center}
    \includegraphics[width=0.48\textwidth]{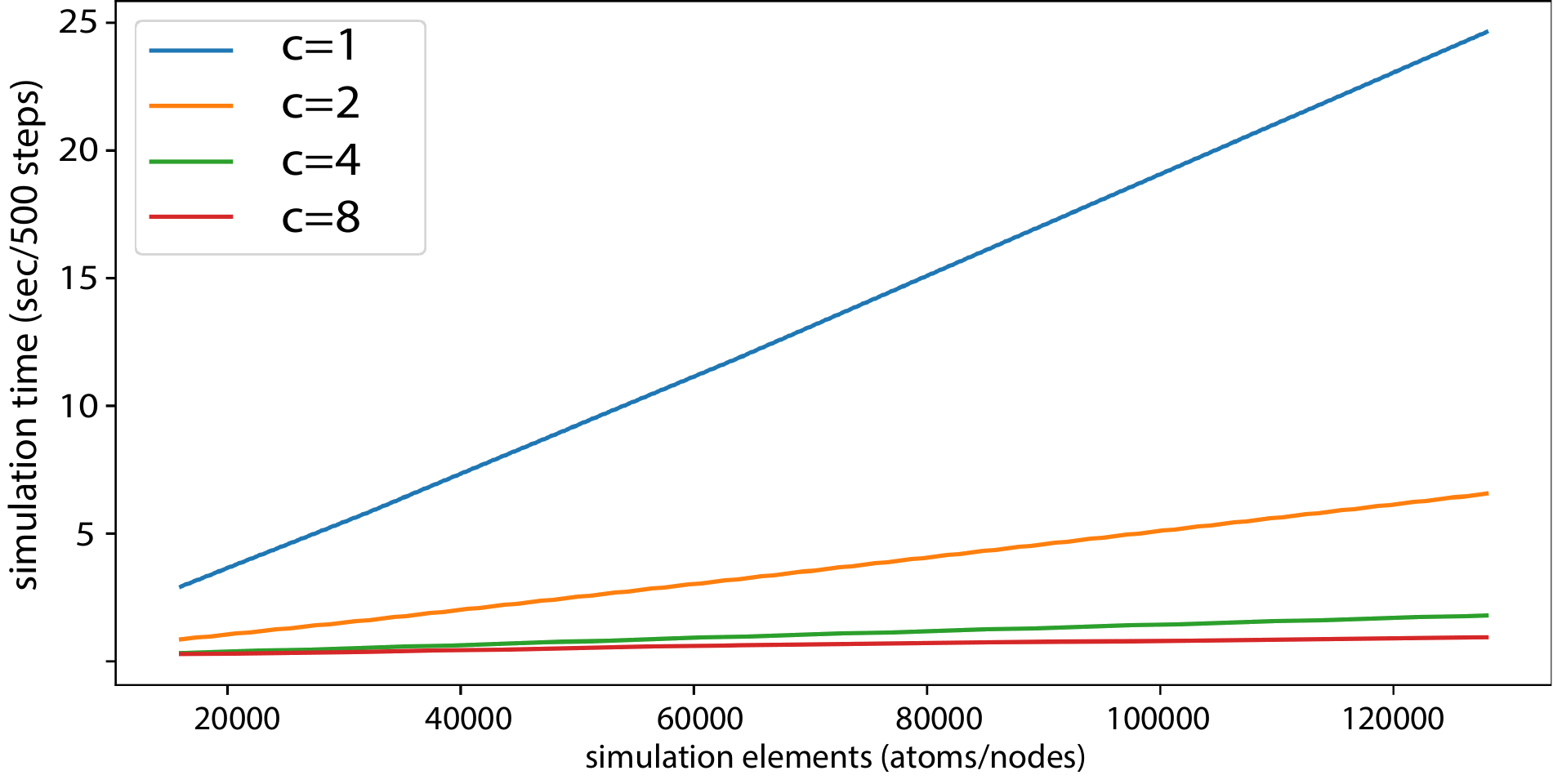}
  \end{center}
  \caption{The dependence of the simulation time on the number of the simulation elements (atoms and nodes) of a 2D ferromagnet with DM interaction on a square lattice. The parameter $c$ (coarseness) represents the number of atoms that fit between the continuum mesh nodes. The time step of \SI{1}{fs} is used.}
  \label{fig3}
\end{figure}

In the above simulations, the introduction of micromagnetic regions brings an improvement of computational efficiency, compared to purely atomistic simulations. To characterise the coarseness of the system, an additional variable $c$ is introduced, which represents the number of atoms that fit between the continuum mesh nodes. Note that this parameter only specifies the coarseness of the simulation nodes of the continuum region. Hence, the value $c=1$ specifies the case when the atomistic and continuum regions have equal density of simulation atoms and simulation nodes, a case which represents essentially the computational effort of a purely atomistic simulation. In Fig.~\ref{fig3}, the simulation time (cpu clock time measured in seconds after the simulation reaches $500$ time steps) is shown for a 2D ferromagnet with rectangular shape, in which both Heisenberg exchange and DM interaction are significant, as a function of the number of simulation elements (atoms/nodes), for different values of $c$, i.e. for different coarseness. The material parameters of this simulation are given in Supplementary Note S3c. The geometry considered in the simulations had rectangular shaped regions of continuum and atomstic regions.
In the simulation considered here, the initial magnetisation points along the $z$-direction and since there is no external magnetic field, the magnetisation evolves in time towards a spin-spiral state. 
The simulation time scales linearly with the size of the system for any choice of coarseness, as shown in Fig.~\ref{fig3}. It may also be seen that the simulation time decreases significantly with increasing values of $c$. Naturally this comes at a cost, in terms of accuracy, and a suitable choice of $c$ must always be chosen with care, depending on shape of the simulation cell and the dynamical responce one is interested in. 
A closer inspection of the computational efficiency of the multiscale approach (data not shown) shows that the efficiency scales exponentially with $c$, clearly demonstrating the major advantage of the multi-scale modeling approach suggested here, compared to a stright atomistic description.




\subsubsection{Single-skyrmion creation via microwave fields} 

\begin{figure*}[tb]
  \begin{center}
    \includegraphics[width=\textwidth]{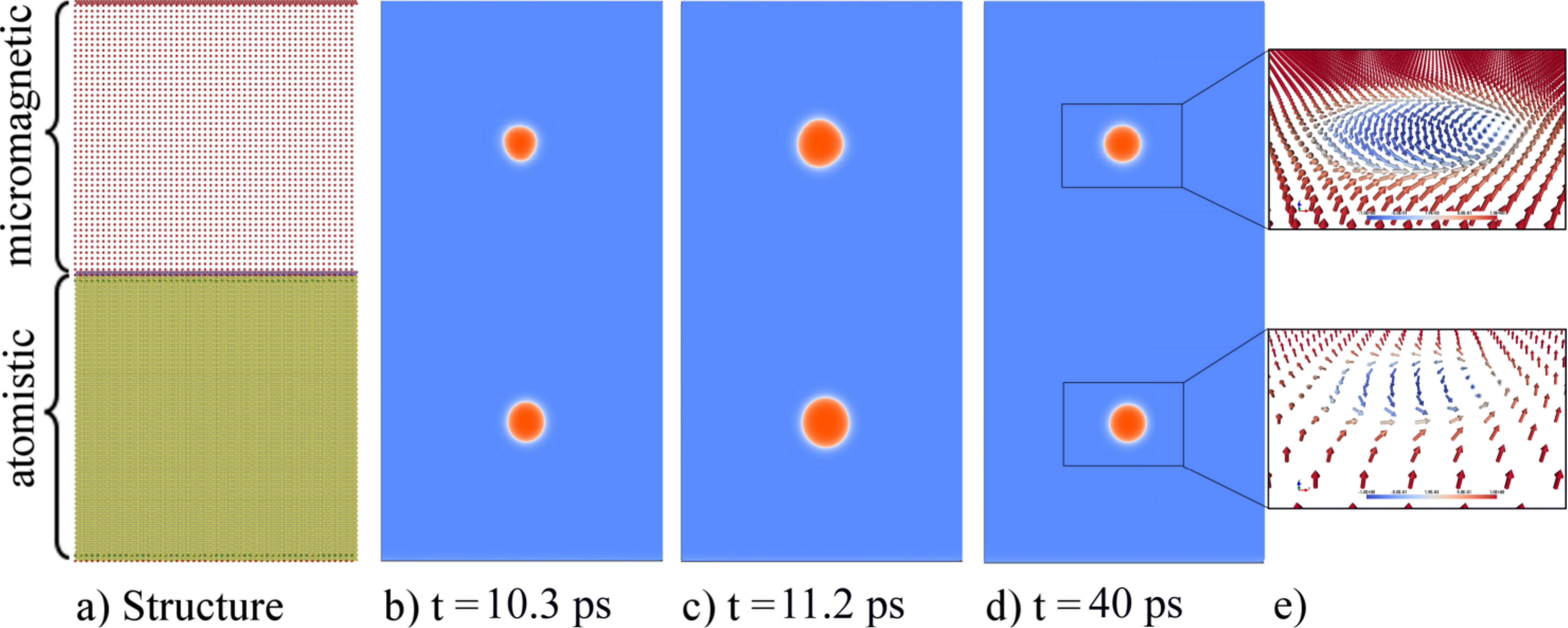}
  \end{center}
  \caption{Skyrmion introduced in the atomistic and the continuum regions of the same material. The size and the shape are shown as functions of time. The sizes of the atomistic and the micromagnetic regions are indicated on the left hand side of the figure. On the right panels, the inner magnetic structures of the vortex skyrmions are shown.}
  \label{fig4}
\end{figure*}

In this section, the creation process of skyrmions via an external magnetic field is illustrated, in a magnetic medium in which both the Heisenberg exchange and the DM interaction are important. Simulation details and materials parameters are given in Supplementary Note S3d). Initially, the magnetic moments point along the $z$-direction and are forced to align parallel by an external field of \SI{6}{T}. Single skyrmions are then stabilized by a local torque. Details are given in Supplementary Note S5. The simulated time evolution of these skyrmions is shown in Fig.~\ref{fig4}. The results from the micromagnetic model are shown in the topmost region, while the atomistic approach, on a hexagonal crystal lattice, is presented in the bottom region. The external field is applied to the entire material throughout the simulations, as it is found to keep the majority of magnetisation aligned along the initial direction, which bounds the size of the skyrmion \cite{hoffmann}.

The skyrmions shown in Fig.~\ref{fig4} represent a metastable state. In absence of the external field, the system can lower its energy by expanding the radius of the skyrmion, while due to the external field, atomic moments favour alignment along the $z$-direction. The competition between the interactions results in a skyrmion with a size following a damped oscillatory mode. This behaviour is known as the so-called ``breathing mode''. In Fig.~\ref{fig4} the inner magnetic texture of the skyrmions is shown, both for the atomistic and the micromagnetic regions. Due to the selected direction of the DM vector, the topological excitations in both regions are vortex skyrmions as shown in Fig.~\ref{fig4}e) and it can be seen that they have similar shape and size \cite{note1}. The breathing mode is responsible for the alternating radius of the skyrmions depicted for various time steps in the simulations as shown in Fig.~\ref{fig4}b)-d). Since the resolution of the pulsed local field is constrained by the resolution of the underlying domain, slight differences in its effect are expected and are revealed at the first few steps of the simulation. This illustrates that a continuum description, even when tuned to have materials parameters selected to be as close as possible to represent the behaviour of an atomistic region, does not capture the correct time evolution of the more accurate atomistic model. However, as the magnetic configuration evolves towards a stable stationary state, skyrmions of the continuum and atomistic descriptions become increasingly similar as shown in Fig.~\ref{fig4}d). The time evolution of both skyrmions is shown in Supplementary Video 5.

\subsubsection{Skyrmion-micro-stress interaction} 

\begin{figure}[tb]
  \begin{center}
    \includegraphics[width=0.48\textwidth]{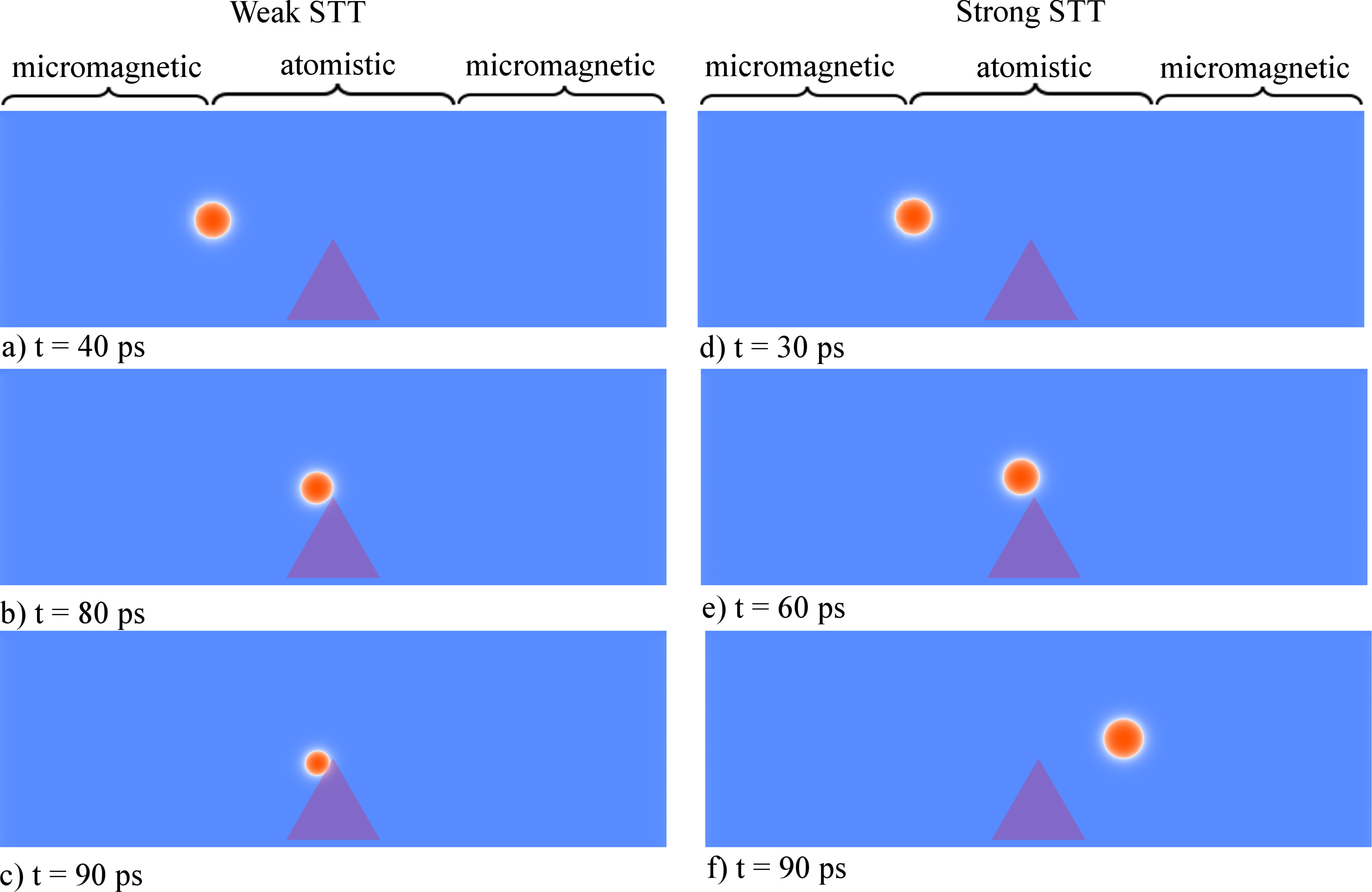}
  \end{center}
  \caption{Skyrmion interaction with the miro-stress-caused defect for two different STT strength values, after inducing a skyrmion via a microwave field. The triangular region marks the area where the magnetic anisotropy is larger than in the other regions. The sizes of the atomistic and the micromagnetic regions are indicated at the top of the figure.}
  \label{fig5}
\end{figure}

In this example, a smaller region of different magnetic anisotropy is considered, where locally residual micro-stresses in a material causes an enhanced anisotropy constant \cite{pofm_dw}. The simulation shown in Fig.~\ref{fig5} is set up similarly to that of Ref.~\cite{fert2013skyrmions}; however, the triangular region with unique anisotropy (coloured differently in the figure) is treated by an atomistic model on a hexagonal lattice. A hedgehog skyrmion is introduced (created as described in the previous section) and driven along the track by means of a spin current induced by STT. The material parameters for this example are given in Supplementary Note S3e). The direction of the spin current used in this simulation is given by vector $(0.995,0.1,0)$. Small $y$-component is used to compensate the Magnus force that drives the skyrmion downwards as already reported in Ref.~\cite{iwasaki2013current}, resulting in motion along a straight line. A weak spin-current of \SI{4.5}{\meter\per\second} was found to be unable to push the skyrmion past the region of enhanced anisotropy. In this case, the skyrmion moves towards the irregularity, where it dissipates energy and angular momentum, which leads to the reduction in size, as shown in Fig.~\ref{fig5}a)-c) and Supplementary Video 6. A strong spin-current induced by the STT can push the skyrmion past the region of enhanced anisotropy --- a value of \SI{6}{\meter\per\second} is used for the simulation in Fig.~\ref{fig5}d)-f) and Supplementary Video 7 and was found to be strong enough to push the skyrmion past the micro-stress region. An intermediate spin-current of \SI{5}{\meter\per\second} sustains the skyrmion as a stationary entity, even when it meets the magnetic irregularity (data not shown).

\subsubsection{Skyrmion-dislocation interaction} 

\begin{figure}[tb]
  \begin{center}
    \includegraphics[width=0.48\textwidth]{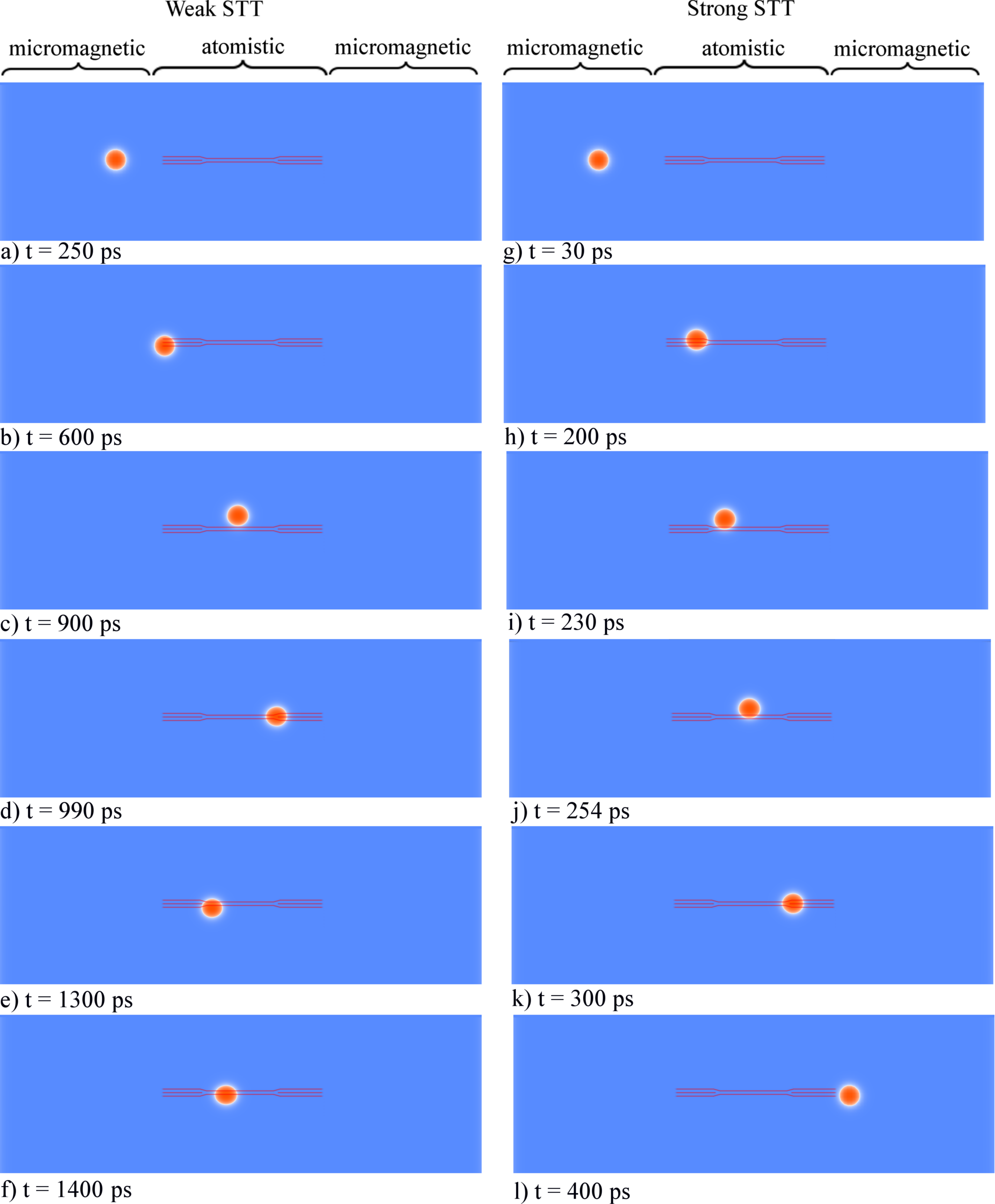}
  \end{center}
  \caption{Skyrmion-dislocation interaction, illustrated at different snapshots in time. The red lines represent atomic planes and visualize the dislocation. The brackets on the top of the figure indicate the atomistic and the micromagnetic regions. The dislocation is fully included in the atomistic region.}
  \label{fig6}
\end{figure}

An edge dislocation is a common crystallographic defect where one plane of atoms is missing a half, which distorts nearby planes of atoms. Such defects are unavoidable in experimental samples, where they may influence the magnetic properties in a decisive way and in this subsection, the dislocation-affected skyrmion motion is illustrated. Results of this section cannot be obtained by a continuum model alone and illustrate the strength of the atomistic-continuum multi-scale approach. The process of determining the positions of atoms within the dislocation is described in the Supplementary Note S6. Material parameters are given in Supplementary Note S3f). 

The sequence of images in Fig.~\ref{fig6} shows snapshots of the skyrmion when it moves across the computational domain. Fig.~\ref{fig6}a)-f), Supplementary Video 8 and Fig.~\ref{fig6}g)-l), Supplementary Video 9 correspond to the STT vectors with components $(1.25,0.175,0)$ and $(5,0.5,0)$ \SI{}{\metre\per\second} respectively. In the latter case (i.e. relatively strong STT), the skyrmion interacts with the defect, however, it overcomes the region around the defect and subsequently leaves the atomistic region, moving to the continuum. For a weaker STT current, the skyrmion moves at the top of the row of atoms of the dislocation. Once it comes to the end of the missing row of atoms, which is indicated by the red lines in Fig.~\ref{fig6}, it moves to the bottom of the dislocation and changes the direction --- it moves backwards, against the direction of the STT. After reaching for the second time the left hand side of the dislocation defect, the direction the skyrmion reverses again. Such oscillatory behaviour of the skyrmion around the dislocation is repeated several times. Due to damping, the skyrmion travels a shorter path each iteration, until it is finally trapped by the defect as shown in Fig.~\ref{fig6}f). This counter-intuitive behaviour, where for a certain time the skyrmion moves against the direction favoured by the STT, has not been reported in literature before and is impossible to observe within a continuum model. 

\begin{figure}[tb]
  \begin{center}
    \includegraphics[width=0.48\textwidth]{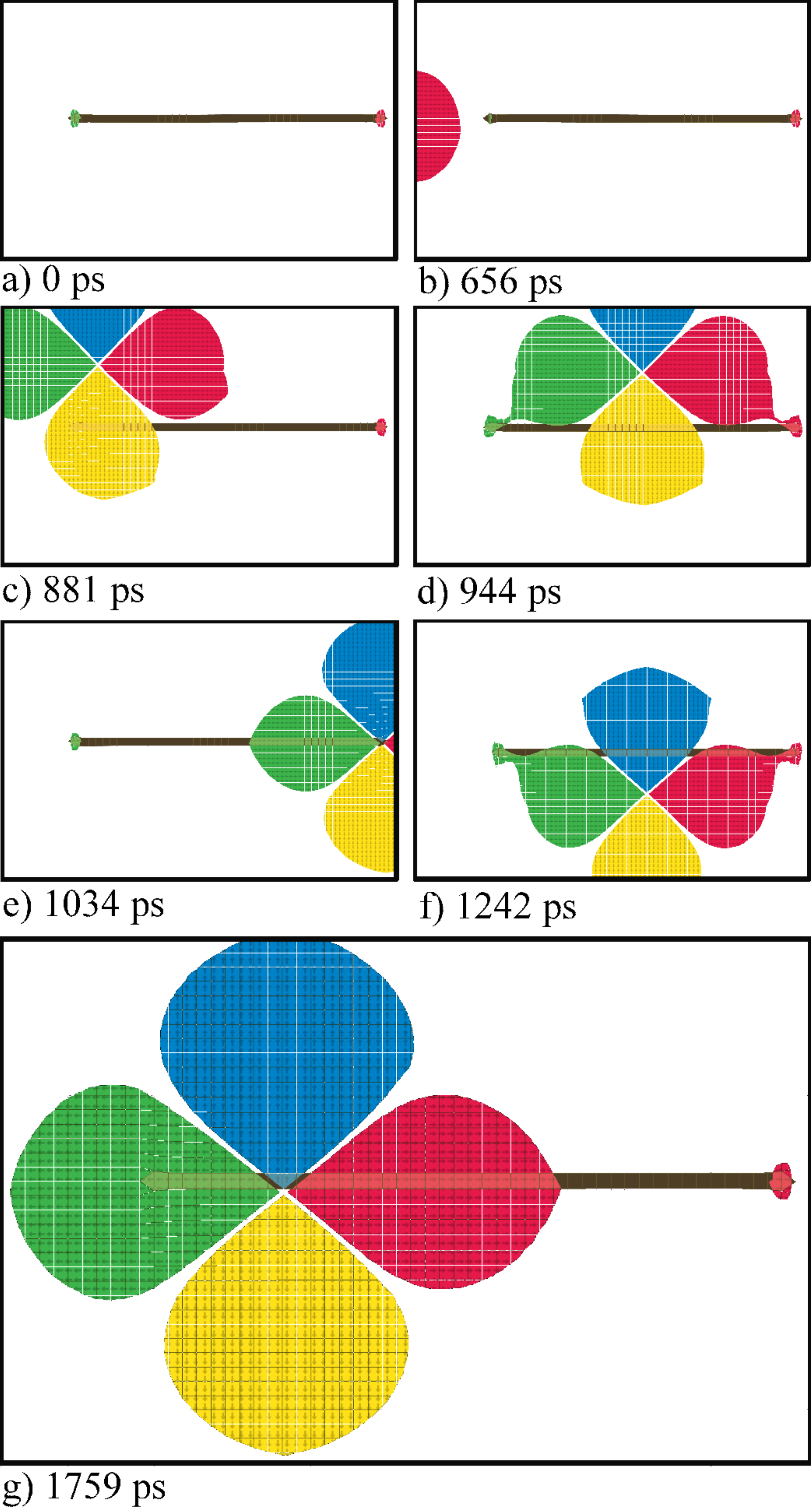}
  \end{center}
  \caption{Skyrmion-dislocation interaction, illustrated at different moments in time, for which only $x$ and $y$ components of the magnetisation are shown. The simulation of Fig.~\ref{fig6}a)-f) is demonstrated. Red, yellow, green and blue colours stand for the directions of the magnetisation along right, down, left and up directions, respectively. The brown/dark line visualizes the dislocation.}
  \label{fig7}
\end{figure}

In Fig.~\ref{fig7} and Supplementary Video 10, the time evolution of $x$ and $y$ components of the magnetisation of the skyrmion is shown in the neighbourhood of the dislocation. The colours indicate the direction of the major component of the magnetisation in the $xy$-plane --- red, yellow, green and blue colours correspond to the directions of the magnetisation primarily along right, down, left and up directions, respectively. The same colour pattern is used to describe the topology of the magnetic interactions at the endpoints of the dislocation. Since the DM vector is in-plane and is perpendicular to the bond direction, the dislocation on the left endpoint introduces a local distortion of the DM interaction in such a way that it favours magnetisation pointing opposite to $x$-direction. For the right endpoint of the dislocation, the situation is reversed. As a consequence, the ground-state magnetic configuration is achieved when the atomic magnetic moments on the left and the right endpoints are tilted to the left and to the right, respectively, as shown in green and red colors in Fig.~\ref{fig7}a). The white background colour indicates that the magnetisation is pointing along the $z$-direction.
  
As depicted in Fig.~\ref{fig7}, the skyrmion stays on the upper side of the defect when moving to the right and in the lower side when moving to the left. Initially, the moments on the left and the right sides of the skyrmion point to the left and to the right, respectively, as can be seen in Fig.~\ref{fig7}a). However, interaction with the dislocation introduces an asymmetry in the components of the magnetisation. When the skyrmion is approaching the dislocation, Fig.~\ref{fig7}b), the red part of the skyrmion is facing the green endpoint of the dislocation. This creates a repulsive potential with two local magnetisations directions that are antiparallel and to lower the exchange interaction, these regions try to avoid each other. Since the STT current is pushing the skyrmion to the right in the figure, this external torque overcomes the repulsive potential and the skyrmion is moved along the upper side of the dislocation. The DM interaction coupled to the lattice defect causes a chirality and in the example shown in Fig.~\ref{fig7} favours a clockwise rotation of the magnetisation. This pushes the skyrmion towards the upper part of the dislocation when it is moving to the right. In Fig.~\ref{fig7}d) the skyrmion is located at a position where it is bounded by ferromagnetic exchange interaction at both sides of the dislocation, causing a local minimum in the energy landscape. However, since the skyrmion has a finite linear momentum produced by the STT, it overcomes this bound state. As the skyrmion continues to move to the right along the dislocation it reaches the endpoint and here it stays for a short time on the right side of the dislocation. At this point the STT balance out the intrinsic exchange and DM interaction between the skyrmion and the lattice dislocation, thus, the right part of the skyrmion is pinned at the right side of the dislocation in another bound state, Fig.~\ref{fig7}e). The inertia of the skyrmion together with the clockwise rotation of the magnetisation induced by the DM interaction now favours the skyrmion to move along the left direction under the dislocation. Notice also that the intensity of the spin current induced by the STT is not strong enough to overcome the inertia of the skyrmion during this short time. Figure~\ref{fig7}f) shows another possible bound state, similar to the one in Fig.~\ref{fig7}d), which again is overcome by the inertia of the skyrmion. Finally, in Fig.~\ref{fig7}g) the skyrmion is pinned or trapped by the attractive potential between the green region of the dislocation and the green part of the skyrmion. On this way back to the final state, the skyrmion looses energy and linear momentum until it is trapped in the left side of the dislocation and, since the STT current is not sufficient to take the skyrmion out of this bounding potential, the skyrmion is pinned on the left side of the dislocation.

\section*{Concluding remarks}

In this communication, an efficient computational methodology, which enables a multi-scale description of magnetisation dynamics, and software are presented. The method proposed here is of the partitioned-domain type and is implemented in the UppASD software. Using a number of examples, it is demonstrated that the continuum and the atomistic descriptions can be interfaced in a seamless way. As shown here, this enables simulations of magnetic phenomena at atomistic length scales, which are coupled to a micromagnetic description and which allow simulating cells that are of the same length scale as experimental samples.

As illustration of the multi-scale method, several examples of chiral and topological magnetic states are used --- domain walls and skyrmions that interact with local perturbations of atomistic length scale. With this technique, an atomistic description of the Barkhausen effect is obtained. In addition, skyrmion dynamics is investigated. Of the more conspicuous results presented here, one may note the intricate dynamics of a skyrmion interacting with a lattice defect --- an edge dislocation. The counter-intuitive motion, which may go against an external STT, is analysed in detail and is discussed in terms of the magnetic interactions coupled to the topology of the defect and the topology of the skyrmion. 

The continuum regions have certain discretisation freedom compared to the atomistic region, which leads to the major advantage of the multi-scale approach. However, as dynamics of objects, such as skyrmions and domain walls, is modelled, the mesh density of the continuum region should allow accurate resolution of the modelled objects. This would also mean that sometimes, if magnetisation gradient is relatively large (i.e. fields change relatively rapidly), the continuum mesh must be relatively fine, which alleviates the computational advantages of using the continuum regions. One of the possible solutions in this case, is using the dynamic remeshing and/or dynamically moving atomistic/continuum regions. This, however, might require additional resources and increases the complexity of the software.

The method proposed here allows simulating magnetic phenomena in general, e.g. in the field of magnonics or for studies of racetrack memories, with topological or chiral objects. Simulations of such phenomena calls for realistically-sized simulation cells, which only a micromagnetic simulation model allows for. This level of description must be combined with an atomistic description, to treat regions where unavoidable lattice defects influence the magnetisation dynamics, and the method outlined here allows for such a description opening up for simulations of many exciting magnetic phenomena and technologies. 

\section{Data availability}

The source code of the developed tool is freely available upon request to the authors (http://katalog.uu.se/empinfo/?id=N12-216). 

\section{Acknowledgements}

We are grateful for the support from STandUpp, eSSENCE, the Swedish Research Council, the Knut and Alice Wallenberg Foundation, the Foundation for Strategic Research and the Swedish Energy Agency. We acknowledge D. Thonig for fruitful discussions.



\begin{thebibliography}{}

\bibitem{vzutic2004spintronics}
I. {\v{Z}}uti{\'c}, J. Fabian, and S. Das Sarma.
\newblock Spintronics: Fundamentals and applications.
\newblock {\em Reviews of modern physics} {\bf 76}, 323 (2004).

\bibitem{burkard2000spintronics}
G. Burkard, H.-A. Engel, and D. Loss.
\newblock Spintronics and quantum dots for quantum computing and quantum communication.
\newblock {\em Fortschr. Phys.} {\bf 48}, 965 (2000).

\bibitem{beaurepaire}
E. Beaurepaire, J.-C. Merle, A. Daunois, and J.-Y. Bigot
\newblock Ultrafast Spin Dynamics in Ferromagnetic Nickel
\newblock {\em Phys. Rev. Lett.} {\bf 76}, 4250 (1996).

\bibitem{abert}
C.~W. Abert.
\newblock {\em Discrete Mathematical Concepts in Micromagnetic Computations}.
\newblock PhD thesis, Universitat Hamburg, Von-Melle-Park 3, 20146 Hamburg,
  2013.
  
\bibitem{SDbook}
O. Eriksson, A. Bergman, L. Bergqvist and J. Hellsvik, 
\newblock Atomistic spin-dynamics: fundamentals and applications
\newblock {\em Oxford University Press}, 2017.

\bibitem{garcia2006adaptive}
C.~J. Garcia-Cervera and A.~M. Roma.
\newblock Adaptive mesh refinement for micromagnetics simulations.
\newblock {\em IEEE transactions on magnetics} {\bf 42}, 1648 (2006).

\bibitem{kim2017mimetic}
E. Kim and K. Lipnikov.
\newblock The mimetic finite difference method for the landau--lifshitz
  equation.
\newblock {\em Journal of Computational Physics} {\bf 328}, 109 (2017).

\bibitem{tako1997finite}
K.M.~Tako, T. Schrefl, M.A.~Wongsam, and R.W.~Chantrell.
\newblock Finite element micromagnetic simulations with adaptive mesh
  refinement.
\newblock {\em Journal of applied physics} {\bf 81}, 4082 (1997).

\bibitem{GarciaSanchez2005}
F. Garcia-Sanchez, O. Chubykalo-Fesenko, O. Mryasov, R. W. Chantrell, and K. Y. Guslienko.
\newblock Exchange spring structures and coercivity reduction in FePt/FeRh bilayers: A comparison of multiscale and micromagnetic calculations.
\newblock Applied Physics Letters {\bf 87}, 122501 (2005).

\bibitem{Jourdan2008}
T. Jourdan, A. Marty, and F. Lancon.
\newblock Multiscale method for Heisenberg spin simulations.
\newblock Physical Review B {\bf 77}, 224428 (2008).

\bibitem{andreas2014multiscale}
C. Andreas, A. K{\'a}kay, and R. Hertel.
\newblock Multiscale and multimodel simulation of bloch-point dynamics.
\newblock {\em Physical Review B} {\bf 89}, 134403 (2014).

\bibitem{de2016multiscale}
A. De~Lucia, B. Kr{\"u}ger, O.~A. Tretiakov, and M. Kl{\"a}ui.
\newblock Multiscale model approach for magnetization dynamics simulations.
\newblock {\em Phys. Rev. B} {\bf 94}, 184415 (2016).

\bibitem{Hertel2018}
R. Hertel.
\newblock Applications of multi-scale modeling to spin dynamics in spintronics devices.
\newblock In Handbook of Materials Modeling, Springer International Publishing, 2018.

\bibitem{skubic2008method}
B. Skubic, J. Hellsvik, L. Nordstr{\"o}m, and O. Eriksson.
\newblock A method for atomistic spin dynamics simulations: implementation and
  examples.
\newblock {\em Journal of physics: condensed matter} {\bf 20}, 315203 (2008).

\bibitem{webUppasd}
{UppASD} uppsala university project website.
\newblock \url{http://physics.uu.se/uppasd}.

\bibitem{Poluektov2018}
M.~Poluektov, O.~Eriksson, and G.~Kreiss.
\newblock Coupling atomistic and continuum modelling of magnetism.
\newblock {\em Computer Methods in Applied Mechanics and Engineering}
  {\bf 329}, 219 (2018).

\bibitem{Poluektov2016}
M.~Poluektov, O.~Eriksson, and G.~Kreiss.
\newblock {S}cale transitions in magnetisation dynamics.
\newblock {\em Communications in Computational Physics} {\bf 20}, 969 (2016).

\bibitem{Tadmor2011}
E. B. Tadmor and R. E. Miller.
\newblock Modeling materials.
\newblock Cambridge University Press, 2011.

\bibitem{Ortner2014}
C. Ortner and L. Zhang.
\newblock Energy-based atomistic-to-continuum coupling without ghost forces.
\newblock Computer Methods In Applied Mechanics and Engineering {\bf 279}, 29 (2014).

\bibitem{Qu2005}
S. Qu, V. Shastry, W. A. Curtin, R. E. Miller.
\newblock A finite-temperature dynamic coupled atomistic/discrete dislocation method.
\newblock Modelling Simulation Mater. Sci. Eng. {\bf 13}, 1101 (2005).

\bibitem{note}
A multi-domain configuration is in general stabilized only if a dipole interaction is also included in the effective field. We omit this interaction in what follows and replace it by a magnetic structure prepared in a multi-domain state. Since the domain wall structure is governed by the anisotropy and exchange interactions only, this step does not influence the shape or dynamis of the domain wall.

\bibitem{aharoni}
A. Aharoni.
\newblock Introduction to the Theory of Ferromagnetism.
\newblock Oxford Science Publications, 2nd ed., 2000.

\bibitem{chubykalo}
O. A. Chubykalo J. M. Gonzalez J. Gonzalez
\newblock Barkhausen jump distributions in a micromagnetic model
\newblock {\em J. Magn. Magn. Mater.} {\bf 184}, L257 (1998).


\bibitem{pereiro}
M. Pereiro, D. Yudin, J. Chico, C. Etz, O. Eriksson and A. Bergman.
\newblock Topological excitations in a kagome magnet.
\newblock {\em Nat. Commun.} {\bf 5}, 4815 (2014).

\bibitem{pereiro1}
K. Koumpouras, D. Yudin, C. Adelmann, A. Bergman, O. Eriksson and M. Pereiro.
\newblock A majority gate with chiral magnetic solitons.
\newblock {\em J. Phys.: Condens. Matter} {\bf 30}, 375801 (2018).

\bibitem{hoffmann}
W. Jiang, P. Upadhyaya, W. Zhang, G. Yu, M. B. Jungfleisch, F. Y. Fradin, J. E. Pearson, Y. Tserkovnyak, K. L. Wang, O. Heinonen, S. G. E. te Velthuis, and A. Hoffmann.
\newblock Blowing magnetic skyrmion bubbles.
\newblock {\em Science} {\bf 349}, 283 (2015).

\bibitem{heinze2011spontaneous}
S. Heinze, K. Von~Bergmann, M. Menzel, J. Brede, A.
  Kubetzka, R. Wiesendanger, G. Bihlmayer, and S. Bl{\"u}gel.
\newblock Spontaneous atomic-scale magnetic skyrmion lattice in two dimensions.
\newblock {\em Nature Physics} {\bf 7}, 713 (2011).
\bibitem{note1}
It must be mentioned that since the process of skyrmion creation was the same in the atomistic and the continuum regions, within the latter, in the initial phase the skyrmion is not sufficiently continuous, but it rapidly evolves to a `smooth' solution (term `smooth' is not used in strict mathematical sence here). This discrepancy could be the origin of the difference in temporal behaviour. Therefore, if one is interested in the precise initial phase, exclusively atomistic description must be used. However, at the final state, atomistic and continuum skyrmions are equivalent and both descriptions result in a breathing mode.

\bibitem{pofm_dw}
S. Chikazumi and S. H. Charap.
\newblock {\em Physics of Magnetism}.
\newblock Krieger Pub Co, 1978.


\bibitem{fert2013skyrmions}
A. Fert, V. Cros, and J. Sampaio. 
\newblock Skyrmions on the track. 
\newblock {\em Nature nanotechnology} {\bf 8}, 152 (2013).

\bibitem{iwasaki2013current}
J. Iwasaki, M. Mochizuki, and N. Nagaosa.
\newblock Current-induced skyrmion dynamics in constricted geometries.
\newblock {\em Nature nanotechnology} {\bf 8}, 742 (2013).

\bibitem{jwmorris}
J. W.~Morris Jr.
\newblock Materials science.
\newblock
  \url{http://www.mse.berkeley.edu/groups/morris/MSE200/I-structure.pdf}, 2007.

\end{thebibliography}
\end{document}


\title{A multiscale approach for magnetisation dynamics: unraveling exotic magnetic states of matter}

\author{\'{E}. M\'{e}ndez}
\affiliation{Division of Materials theory, Department of Physics and Astronomy, Uppsala University, Box 516, 75121 Uppsala, Sweden}
\author{M. Poluektov}
\affiliation{International Institute for Nanocomposites Manufacturing, WMG, University of Warwick, Coventry CV4 7AL, UK}
\author{G. Kreiss}
\affiliation{Department of Information Technology, Uppsala University, Box 337, SE-751 05 Uppsala, Sweden}
\author{O. Eriksson}
\affiliation{Division of Materials theory, Department of Physics and Astronomy, Uppsala University, Box 516, 75121 Uppsala, Sweden}
\affiliation{School of Science and Technology, \"Orebro University, SE-70182 \"Orebro, Sweden}
\author{M. Pereiro}
\affiliation{Division of Materials theory, Department of Physics and Astronomy, Uppsala University, Box 516, 75121 Uppsala, Sweden}


\begin{abstract}
\center{Contribution from the Division of Materials Theory, Uppsala University. This document is not subject to copyright.}
\end{abstract}

\maketitle

\section{Supplementary note S1}
In the atomistic approach, the dynamic behaviour of magnetic moments of individual atoms is described by the atomistic Landau-Lifshitz-Gilbert equation \cite{Bergqvist2013, Evans2014}:
\begin{equation}
  \frac{\partial}{\partial t} {\bf m}_i = -\beta_\mathrm{L} {\bf m}_i \times \left( {\bf h}_i + b_1 {\bf \tau}_i \right) - \alpha_\mathrm{L} {\bf m}_i
  \times \left( {\bf m}_i \times \left( {\bf h}_i + b_2 {\bf \tau}_i \right) \right),
  \label{eq:ASD_LLG}
\end{equation}
\begin{equation}
  {\bf \tau}_i = {\bf g} \cdot \sum_k {\bf s}_{ik} {\bf m}_k , \quad {\bf s}_{ik} = {\bf x}_k - {\bf x}_i ,
  \label{eq:ASD_STT}
\end{equation}
\begin{equation}
  \beta_\mathrm{L} = \frac{\gamma}{\mu\left(1+\lambda^2\right)} ,
  \quad \alpha_\mathrm{L} = \beta_\mathrm{L}\lambda ,
  \label{eq:ASD_param}
\end{equation}
where $\gamma$ is the gyromagnetic ratio, $\lambda$ is the phenomenological Gilbert damping constant, ${\bf m}_i$ is the direction of spin magnetic moment with $\left|{\bf m}_i\right| = 1$, 
$\mu$ is the length of spin magnetic moment and ${\bf h}_i$ is the effective field. The spin-transfer torque (STT) is represented by the vector ${\bf \tau}_i$ while  $b_1$, $b_2$ and the components of the vector ${\bf g}$ are STT coefficients characteristic of the material. The vector  ${\bf s}_{ik}$ is connecting atoms $i$ and $k$ (where ${\bf x}_k$ are atomic positions) and the summation is over atoms in a neighborhood of atom $i$, such that $k \neq i$. The following form of the effective field is considered:
\begin{equation}
  {\bf h}_i = \left( \sum_j J_{ij} {\bf m}_j \right) + \left( \sum_j {\bf D}_{ij} \times {\bf m}_j \right) + {\bf K}_\mathrm{a} \cdot {\bf m}_i + \mu {\bf h}_\mathrm{e} ,
  \label{eq:ASD_H}
\end{equation}
where $J_{ij}$ are constants of the Heisenberg exchange interaction between atoms $i$ and $j$, vectors ${\bf D}_{ij}$ describe the Dzyaloshinskii-Moriya (DM) interaction, ${\bf K}_\mathrm{a}$ is the anisotropy tensor and ${\bf h}_\mathrm{e}$ is the external field. Summations are taken for all $j$, such that $j \neq i$, where any number of neighbours (i.e. non-nearest) can be included.

At the continuum scale, the magnetisation dynamics is modelled by the continuum version of the Landau-Lifshitz-Gilbert (LLG) equation \cite{Aharoni1996, Cimrak2008}:
\begin{equation}
  \frac{\partial}{\partial t} {\bf m} = -\beta_\mathrm{L} {\bf m} \times \left( {\bf h} + b_1 {\bf \tau} \right) - \alpha_\mathrm{L} {\bf m}
  \times \left( {\bf m} \times \left( {\bf h} + b_2 {\bf \tau} \right) \right),
  \label{eq:C_LLG}
\end{equation}
\begin{equation}
  {\bf \tau} = {\bf d} \cdot \nabla {\bf m} ,
  \label{eq:C_STT}
\end{equation}
where ${\bf m}$ is the normalised magnetisation field ($|{\bf m}| = 1$) and $\beta_\mathrm{L}$ and $\alpha_\mathrm{L}$ constants have the same meaning as in Eq.~(\ref{eq:ASD_LLG}). In the case of the continuum approach, the following effective field is considered:
\begin{equation}
  {\bf h} = {\bf A}_\mathrm{e} : \nabla\nabla {\bf m} + \nabla \cdot \left( {\bf D}_\mathrm{e} \times {\bf m} \right)+ {\bf K}_\mathrm{a} \cdot {\bf m} + \mu {\bf h}_\mathrm{e} ,
  \label{eq:C_H}
\end{equation}
where tensor ${\bf A}_\mathrm{e}$ contains the exchange interaction constants while the  tensor ${\bf D}_\mathrm{e}$ includes the Dzyaloshinskii-Moriya interaction constants. The anisotropy tensor is represented by  ${\bf K}_\mathrm{a}$ and ${\bf h}_\mathrm{e}$ is the external field. After a quick inspection of Eqs.~(\ref{eq:ASD_LLG}) and (\ref{eq:C_LLG}), it is easy to see that both equations retain a similar mathematical form. It is straightforward to prove that both equations are invariant under the following transformations: 
\begin{equation}
  {\bf A}_\mathrm{e} = \frac{1}{2} \sum_{j,j \neq i} J_{ij} {\bf s}_{ij} {\bf s}_{ij} ,
  \label{eq:C_exch_A}
\end{equation}
\begin{equation}
  {\bf D}_\mathrm{e} = \sum_{j, j \neq i} {\bf s}_{ij} {\bf D}_{ij} ,
  \label{eq:C_exch_D}
\end{equation}
\begin{equation}
  {\bf d} = {\bf g} \cdot \sum_{k, k \neq i} {\bf s}_{ik} {\bf s}_{ik} ,
  \label{eq:C_STT_const}
\end{equation}
where ${\bf s}_{ij}$ is the vector connecting atoms $i$ and $j$. Since the anisotropy term is local, the same anisotropy tensor ${\bf K}_\mathrm{a}$ is used in the continuum and the atomistic equations. The tensors ${\bf A}_\mathrm{e}$ and ${\bf D}_\mathrm{e}$ and the vector ${\bf d}$ are spatially homogeneous, i.e. these tensor/vector parameters do not depend on $i$.

The usual way of comparing the atomistic and the continuum models is by considering the continuum solution that coincides with the atomistic solution at all lattice points \cite{Luskin2013}. To find the difference between the models, the continuum solution is fixed and the asymptotic behaviour of the difference is found as a function of the atomistic lattice spacing. In the case of magnetism, a continuum magnetisation field ${\bf m}$ is considered in such a way that it coincides with the atomistic spin magnetic moments ${\bf m}_k$ at lattice positions, i.e. ${\bf m}\left({\bf x}_k\right) = {\bf m}_k$, where ${\bf x}_k$ are atomic positions \cite{Poluektov2018}.

Most atomistic lattices of crystalline materials can be categorised as point-symmetric atomistic lattices, i.e. for each pair of atoms $i$ and $j$, there exists an atom $k$ which is of the same type as atom $j$, such that ${\bf s}_{ij} = - {\bf s}_{ik}$. The derivation of Eqs.~(\ref{eq:C_exch_A}) and (\ref{eq:C_exch_D}) can be found in Ref.~\cite{Poluektov2018}, where it is shown that for the point-symmetric atomistic lattices, the atomistic and the continuum models are consistent with the error estimate
\begin{equation}
  \left|{\bf h}_i - {\bf h}\right| = O\left(a^2\right) ,
\end{equation}
as $a \to 0$, where $a$ is the atomistic lattice spacing and the effective fields are indicated in Eqs.~(\ref{eq:ASD_H}) and (\ref{eq:C_H}).

The proof of the consistency between the atomistic an the continuum spin-transfer torque terms, Eqs.~(\ref{eq:ASD_STT}) and (\ref{eq:C_STT}), respectively, can easily be derived by Taylor-expanding solution ${\bf m}\left({\bf x}_i\right) = {\bf m}_i$, as was performed in Ref.~\cite{Poluektov2018}. This results in the following asymptotic behaviour:
\begin{equation}
  \left|{\bf \tau}_i - {\bf \tau}\right| = O\left(a^2\right) .
\end{equation}

\begin{figure}
  \begin{center}
    \includegraphics[width=0.48\textwidth]{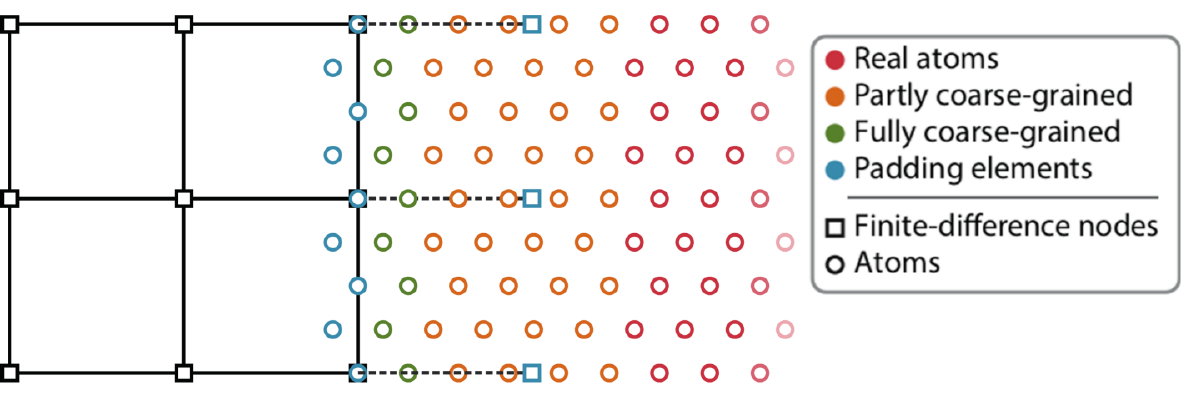}
  \end{center}
  \caption{Schematic representation of the atomistic-continuum interface. The fully atomistic region is represented by real atoms while the fully micromagnetic region is represented by finite-difference nodes. In the interface, both regions mix each other.}
  \label{fig:scheme}
\end{figure}

In multiscale partitioned-domain atomistic-continuum coupling approach, the entire computational region is split into two subregions --- the atomistic and the continuum domains. There is a ``sharp'' atomistic-continuum interface (i.e. a curve in two-dimensional systems, a surface in three-dimensional systems) between these two regions, as illustrated in Fig.~\ref{fig:scheme}.

The continuum region is discretised using the finite-difference method. The atomistic-continuum interface encapsulates a layer of finite-difference nodes and is linear between the neighbouring nodes. The coupling of the regions is implemented by constructing padding atoms and padding nodes, which provide boundary conditions for the atomistic and the continuum regions, respectively. Real atoms interact with padding atoms, while the solution at padding atoms is obtained from the continuum region. Finite-difference nodes interact with padding nodes, while the solution at padding nodes is obtained from the atomistic region. 

The solution at padding atoms is obtained by bilinear interpolation of the continuum solution with normalisation. The normalisation is introduced to preserve the length of spin magnetic moments.
The solution at padding nodes is obtained by normalised weighted average of the atomistic solution inside the box with side $\Delta x$ centred at the padding node, where $\Delta x$ is the inter-node distance. For all atoms inside the box, the weight is assigned as the area of the intersection of the box with side $a$ centred at the atom and the box with side $\Delta x$ centred at the padding node. The normalisation is introduced to preserve the nodal length of the vector field solution.

To reduce the ghost-torque error for systems with non-nearest-neighbour atomistic interactions, the behaviour of atoms close to the atomistic-continuum interface is modified \cite{Poluektov2018}. More specifically, $J_{ij}$ and ${\bf D}_{ij}$ are modified for a finite set of atoms. Two types of modified atoms can be distinguished --- the fully coarse-grained atoms, which interact only with the nearest neighbours, and the partly coarse-grained atoms, which also interact with non-nearest neighbours. The modification of $J_{ij}$ and ${\bf D}_{ij}$ is performed such that the fully coarse-grained atoms completely isolate the atomistic-continuum interface. More details about the exact way of modifying the atoms can be found in Ref.~\cite{Poluektov2018}. 

To reduce the high-frequency wave-reflection from the atomistic-continuum interface, additional numerical damping is added to atoms close to the atomistic-continuum interface \cite{Poluektov2016, Poluektov2018}. This damping acts as a low-pass filter for the waves travelling from the atomistic region to the continuum, as the solution is ``attenuated'' to an average solution within a certain window. Due to dispersive nature of the spin waves, the damping is non-linear and depends on time derivative of the solution. The analysis of the dynamics of the damping layer and the exact form of the modification can be found in \cite{Poluektov2016}. 

The time stepping was performed simultaneously for all degrees of freedom, i.e. all finite-difference nodes and atoms. The solution at padding atoms/nodes at a certain time step is obtained from the continuum/atomistic solution of the same time step. Mid-point \cite{dAquino2005} and Depondt \cite{depondt2009spin} methods were used to solve Eqs.~(\ref{eq:ASD_LLG}) and (\ref{eq:C_LLG}) in time. Some limitions of the multiscale modelling are discussed in Supplementary note S2.

\section{Supplementary note S2}

\noindent{\bf a) Problematic exchange configurations.}\\
  There exist some configurations of Heisenberg or Dzyaloshinskii-Moriya interactions that cannot be represented by the presented micromagnetism model. There are two cases known to the authors. The first example occurs in magnetic materials with large angle between adjacent magnetic atoms, e.g. in antiferromagnets or in non-collinear materials. Examples of such materials are bcc Cr, fcc Fe and most elemental forms of Mn. The problem of these materials is that spins tend to vary abruptly in space. These variations lead to larger errors in differentiation via finite difference. When the moment on consequent nodes is very close to be antiparallel, the result of the differentiation is dominated by numerical error, and the result of such simulation is completely meaningless. Furthermore, the local averages used for padding elements and damping band yield also erroneous results on antiparallel or nearly-antiparallel configurations. Variations of the finite difference method that accept antiferromagnets are available in literature.

  The other class of exchange configurations that are problematic are those for which tensor ${\bf A}_e$ or ${\bf D}_e$  result in a null matrix. It is mathematically possible for this to happen on the Heisenberg exchange, but to the best knowledge of the authors there is no physical material that can exhibit such configuration. A known physical example of this happening for DM interaction is found in the pyrochlore antiferromagnet \cite{elhajal2005ordering}. The planes perpendicular to the [111] direction in this material present a Kagome structure with a DM configuration such that if ${\bf  r}_{ij} = -{\bf  r}_{ik}$ then ${\bf  D}_{ij} = {\bf  D}_{ik}$. This can be achieved in Kagome structures without breaking the antisymmetry of the DM interaction because there is no translational symmetry on the exchange vectors. A consequence of this configuration is that ${\bf  r}_{ij} \; {\bf  D}_{ij} = -{\bf  r}_{ik} \; {\bf  D}_{ik}$. Since in the structure for each neighbour to an atom there exist a neighbour in the opposite direction, it follows that

  \begin{equation}
    \sum_{i \ne j}{{\bf  r}_{ij} \oprod {\bf  D}_{ij}} = 0 = D\!e,
  \end{equation}

  Having a micromagnetics domain with 0 antisymmetric exchange does not represent the effects observable on the atomistic material. As an example, skyrmions and anti-skyrmions can exist on the atomistic representation, but cannot occur on a material without DM interaction.
  
\vspace{0.5cm}
{\bf b) Derivative errors on the interface.}\\

\begin{figure}
  \begin{center}
    \includegraphics[width=0.48\textwidth]{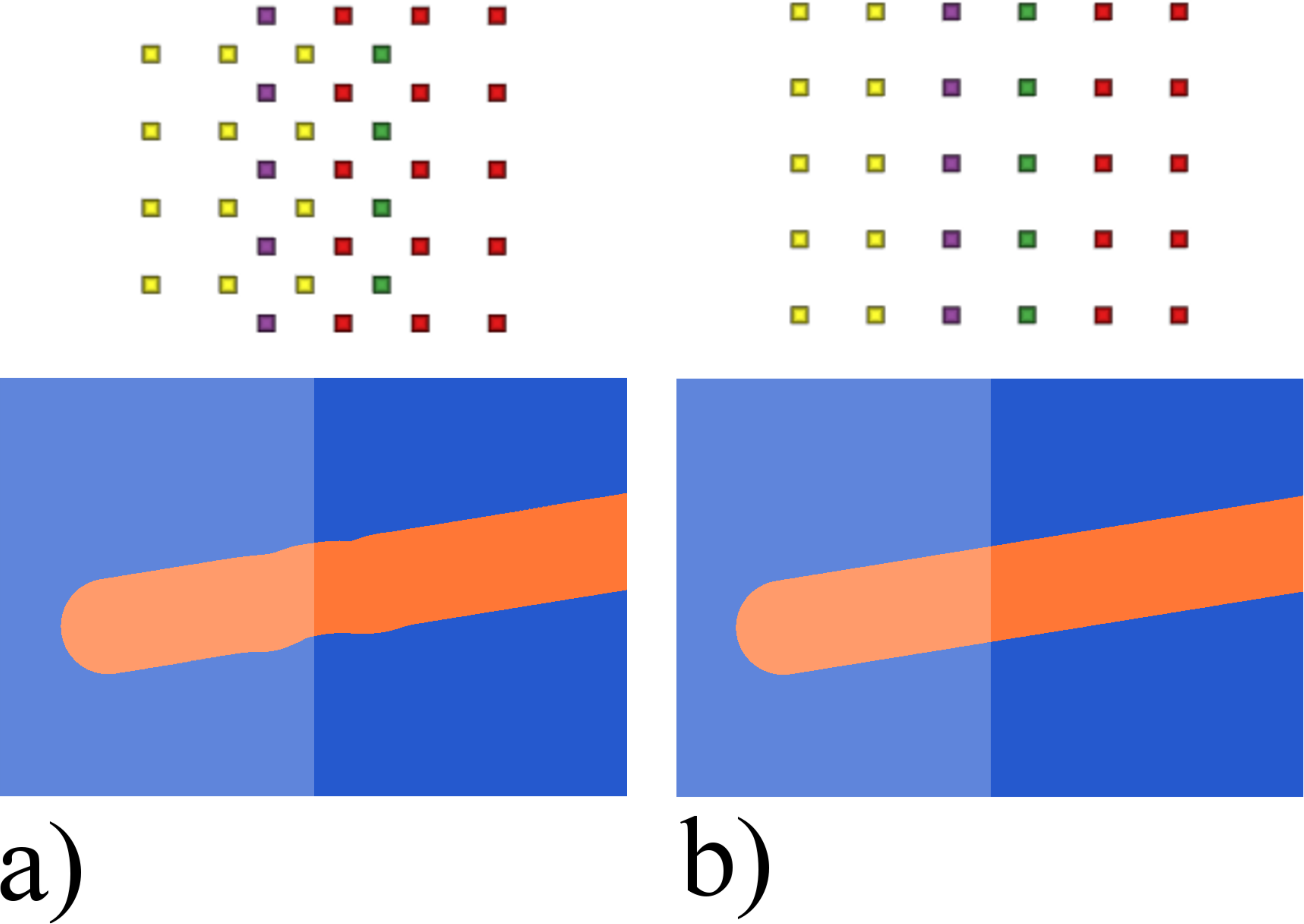}
  \end{center}    
  \caption{
    Trajectory of a skyrmion crossing the boundary.
    Atomistic regions have a white overlay on the left side.    
    (a) The result from an unaligned grid, where the padding node is interpolated from its 4 neighbouring atoms.    
    Distortions can be appreciated on the interfaces.    
    (b) Similar setup with nodes aligned to atoms.
    Padding nodes receive its moment from the atoms at the same positions.
    The distortion does not appear in this case.
  }
  \label{fig1}
\end{figure}

The spatial derivative of the magnetization function is considered in several parts of this model. The micromagnetics effective field terms for exchange indirectly introduce these derivatives via the transformed parameters explained in Supplementary note S4. The spin-transfer torque model directly computes this derivative via finite-differences in the continuum.

When the setup is made in such a way that padding nodes fall very closely or directly over atoms, the error due to interpolation is minimized. As a demonstration, two configurations with $\frac{\Delta x}{a} = 1$, STT, Heisenberg and DM are displayed. On one of them, atoms and nodes are aligned, while on the other, there is an offset of $\frac{a}2$. A skyrmion is introduced via microwave fields and pushed through the boundary. The structure and trajectory of the skyrmion is shown for each setup in Fig.~\ref{fig1}. Thus, in Fig.~\ref{fig1}a), the non-aligned atoms on the interface are shown above. At the bottom of the figure, the trajectory of the skyrmion shows a clear deformation when crossing the interface. Notice that the blue area on the right represents the continuum region. In Fig.~\ref{fig1}b) the aligned atomistic and continuum grids show no deformation on the interface.

When derivatives are computed on nodes next to the interface, the interpolation error from the padding node propagates to the approximated derivative.
\begin{figure}
  \begin{center}
  \includegraphics[width=0.68\textwidth]{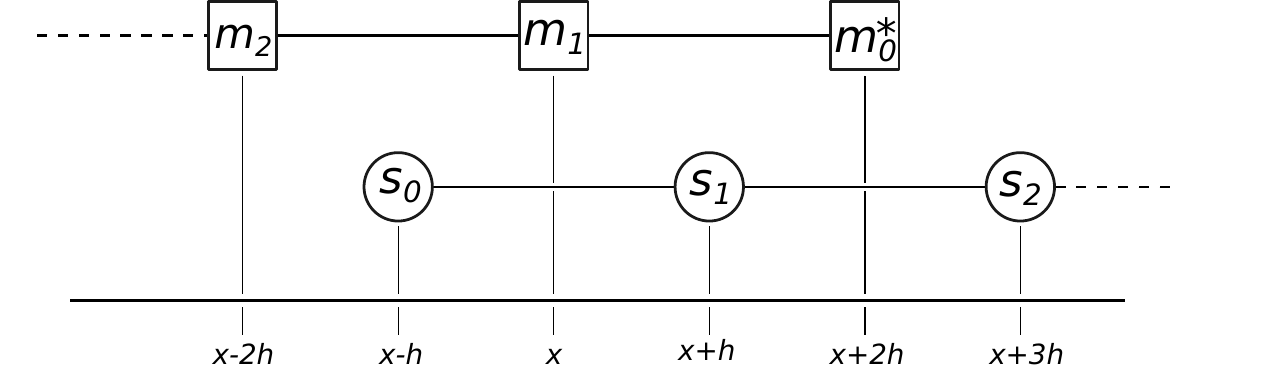}
   \end{center}
  \caption{
    Atoms and nodes at the interface, and positions around ${\bf  m}_1$.
  }
  \label{fig2}
\end{figure}
This error is studied for the 1D setup and shown in Fig.~\ref{fig2}. The elements ${\bf  m}_2$, ${\bf  m}_1$ and ${\bf  m}^\star_0$ are finite difference nodes, while ${\bf  s}_2$, ${\bf  s}_1$ and ${\bf  s}^\star_0$ represent atoms. Elements ${\bf  m}^\star_0$ and $s^\star_0$ are padding elements. The error is studied for the derivative at $m_1$. An exact solution for the magnetization at a point $f(x)$ is assumed to exist. Furthermore, the solution at non-padding elements is assumed to be exact, that is

\begin{align}
  \vec m_2 &= \vec f(x-2h) \\
  \vec m_1 &= \vec f(x)    \\
  \vec s_1 &= \vec f(x+h)  \\
  \vec s_2 &= \vec f(x+3h).
\end{align}
The padding node $m^\star_0$ has its moment determined as
\begin{align}
  \vec m^\star_0 =
  \frac{\vec f(x+h) + \vec f(x+3h)}{\|\vec f(x+h) + \vec f(x+3h)\|},
\end{align}
which, since the moments in our model are always unit vectors, it can be simplified using the cosine law as
\begin{align}
  \vec m^\star_0 =
  \frac{\vec f(x+h) + \vec f(x+3h)}{\sqrt{2+2\cos\theta}} \label{dererr:pre_sim}
\end{align}
where $\theta$ is the angle between $\vec f(x+h)$ and $\vec f(x+3h)$.
When $f(x)$ is sufficiently smooth, the angle $\theta$ is close to 0.
It is known that $f(x)$ needs to be a smooth function to be represented by the continuum, so this assumption does not add a new constrain to the model.
\begin{align}  
  \vec m^\star_0 \approx \frac{\vec f(x+h) + \vec f(x+3h)}{2} \label{dererr:post_sim}
\end{align}
The derivative at $\vec m_1$ is calculated via finite difference as
\begin{align}
  \vec m'_1 = \frac{m^\star_0 - m_2}{4h},
\end{align}
which can be expressed in terms of $f(x)$ as
\begin{align}
  f'(x) \approx \vec m'_1 = 
  \frac{f(x+h)}{8h} + \frac{f(x+3h)}{8h} - \frac{f(x-2h)}{4h}.
\end{align}
Applying Taylor expansion on the previous expression results in
\begin{align}
  \frac{f(x+h)}{8h} &= \frac{1}{8h}f(x) + \frac18f'(x)+\frac{h}{16}f''(x)+... \\
  \frac{f(x+3h)}{8h} &= \frac{1}{8h}f(x) + \frac38f'(x)+\frac{9h}{16}f''(x)+... \\ 
  \frac{-f(x-2h)}{4h} &= \frac{-1}{4h}f(x) + \frac12f'(x)-\frac{8h}{16}f''(x)+... \\ 
  \vec m'_1 &= f'(x)-\frac{5h}{8}f''(x)+... .
\end{align}

This shows the error introduced by this approximation is linear with respect to $h$.
The absolute error is generally small.
It is hard to measure its effect in domain walls.
However, more sensitive configurations, like spin ices or skyrmions, specially when all STT, DM and Heisenberg exchange are present, do show visible artifacts near the interface.
A higher order interpolation is not generally possible since there are not necessarily enough atoms surrounding a padding node.
One possible solution is to discard padding nodes and calculate a finite difference stencil coefficients for each node next to an atom in the interface.
These stencils can be used instead of the suggested in Supplementary note S4 to calculate new exchange parameters for nodes in the interface.

\section{Supplementary note S3}

{\bf Simulation parameters.}
To allow replicating the examples done in this work, the following sections detail the parameters used.
  In all examples, the $x$ coordinate corresponds to the horizontal axis,
  increasing towards the right.
  For 2D examples, the $y$ coordinate corresponds to the vertical axis,
  increasing up, and the $z$ coordinate is out-of-plane, incresing towards the reader.
  
\vspace{0.5cm}
\noindent{{\bf a) Domain walls.}}\\
  \begin{tabular}{ |l|c|  }    
    \hline
    Parameter & Value \\
    \hline 
    Moment magnitude & \SI{2.23}{\mu_B} \\
    Temperature & \SI{0.1}{\milli K} \\
    Damping & 0.5 \\
    $\Delta{x}/a$ & 8 \\
    time-step & \SI{5}{\femto s} \\
    Periodicity & Y axis  \\
    \hline 
    \multicolumn{2}{|l|}{Exchange} \\
    \hline 
    Heisenberg ($J_{ij}$)& \SI{0.6}{\milli Ry} \\
    \hline 
    \multicolumn{2}{|l|}{Anisotropy} \\
    \hline 
    K1 & \SI{-0.0005}{\milli Ry} \\
    K2 & \SI{0.0}{\milli Ry} \\
    direction & $(0,0,1)$ \\
    \hline
    \multicolumn{2}{|l|}{After thermalization} \\
    \hline 
    Field & {$(0,0,-0.2)$}{T} \\    
    \hline
  \end{tabular}
\\
The uniaxial anisotropy favours magnetism along the $z$ direction. 
An external magnetic field of $-0.25$~\SI{}{T} along the $ \vec z$-direction, provides a torque that moves the wall. For the domain wall-stress interaction, the stress presents anisotropy $K1 = \SI{+0.001}{\milli Ry}$ within the defect. 
\\

\vspace{0.5cm}
\noindent{\bf b) Iron-Iridium monolayer.}\\
The parameters for these set of simulations are taken from literature, corresponding to an iron monolayer on iridium\cite{heinze2011spontaneous}.
The parameters are written in function of the magnetic moment magnitude, a moment magnitude of 1 $\mu_B$ is used for simplicity.

For skyrmion-stress another value of DM was used, to tune the size of the skyrmion allowing for larger $\frac{\Delta x}{a}$.
\\
\noindent
  \begin{tabular}{ |l|c|  }    
    \hline
    Parameter & Value \\
    \hline 
    Moment magnitude & \SI{1.0}{\mu_B} \\
    Temperature & \SI{0.1}{K} \\
    Damping & 0.05 \\
    a & \SI{2.715}{\angstrom} \\
    $\Delta{x}/a$ & 2  \\    
    time-step & \SI{1}{\femto s} \\
    Periodicity & X and Y axes  \\
    Crystalline structure & hexagonal \\
    \hline 
    \multicolumn{2}{|l|}{Exchange} \\
    \hline 
    Heisenberg & \SI{0.4189}{\milli Ry} \\
    DM magnitude & \SI{0.1323}{\milli Ry} \\
    DM direction & in-bound  \\
    \hline 
    \multicolumn{2}{|l|}{Anisotropy} \\
    \hline 
    K1 & \SI{-0.05879}{\milli Ry} \\
    K2 & \SI{0.0}{\milli Ry} \\
    direction & $(0,0,1)$  \\
    \hline
    \multicolumn{2}{|l|}{For skyrmion examples.} \\
    \hline 
    DM magnitude & \SI{0.066}{\milli Ry} \\
    \hline
  \end{tabular}
\\
\vspace{0.5cm}

\noindent{\bf c) 2D Ferromagnet with DM interaction.}\\
\\
\noindent
  \begin{tabular}{ |l|c|  }    
    \hline
    Parameter & Value \\
    \hline 
    Moment magnitude & \SI{1.0}{\mu_B} \\
    Temperature & \SI{0}{K} \\
    Damping & 0.1 \\
    a & \SI{1.0}{\angstrom} \\
    time-step & \SI{1}{\femto s} \\
    Crystalline structure & cubic \\
    \hline 
    \multicolumn{2}{|l|}{Exchange} \\
    \hline 
    Heisenberg & \SI{1}{\milli Ry} \\
    DM magnitude & \SI{1}{\milli Ry} \\
    DM direction & \ang{90} wrt bound  \\
    \hline 
  \end{tabular}
\\

\vspace{0.5cm}
\noindent{\bf d) Single skyrmion creation by a microwave field.}\\
A D/J ratio of 30\% was chosen as a realistic value.
J and D are arbitrary, but in the order of what would be expected on a real material.
\\
\noindent

\begin{tabular}{ |l|c|  }    
    \hline
    Parameter & Value  \\
    \hline 
    Moment magnitude & \SI{1.00}{\mu_B} \\
    Temperature & \SI{0.0}{\kelvin} \\
    Damping & 0.05  \\
    a & \SI{2.715}{\angstrom} \\
    $\Delta{x}/a$ & 2  \\    
    time-step & \SI{1}{\femto s} \\
    Periodicity & X and Y axis \\
    Crystalline structure & Hexagonal (atomistic) \\
                          & square (continuum)\\
    \hline 
    \multicolumn{2}{|l|}{Exchange} \\
    \hline 
    Heisenberg &  \SI{0.42}{\milli Ry} \\
    DM magnitude & \SI{0.13}{\milli Ry} \\
    DM direction & Along the bound \\
    \hline 
    \multicolumn{2}{|l|}{Anisotropy} \\
    \hline 
    K1 & \SI{-0.059}{\milli Ry} \\
    K2 & \SI{0.0}{\milli Ry} \\
    direction & $(0,0,1)$  \\
    \hline
  \end{tabular}
\\

\vspace{0.5cm}
\noindent{\bf e) Skyrmion-micro-stress interaction.}\\
A D/J ratio of about 20\% was chosen as a realistic value.
J and D are arbitrary, but in the order of what would be expected on a real material.
\\
\noindent

\begin{tabular}{ |l|c|  }    
    \hline
    Parameter & Value  \\
    \hline 
    Moment magnitude & \SI{1}{\mu_B} \\
    Temperature & \SI{1}{\milli\kelvin} \\
    Damping & 0.1  \\
    a & \SI{1}{\angstrom} \\
    $\Delta{x}/a$ & 2  \\    
    time-step & \SI{1}{\femto s} \\
    Periodicity & Y axis \\
    Crystalline structure & square \\
    \hline 
    \multicolumn{2}{|l|}{Exchange} \\
    \hline 
    Heisenberg &  \SI{0.418}{\milli Ry} \\
    DM magnitude & \SI{0.066}{\milli Ry} \\
    DM direction & \ang{90} wrt bound \\
    \hline 
    \multicolumn{2}{|l|}{Anisotropy} \\
    \hline
    Triangular region &\\
    \hline 
    K1 & \SI{-0.11}{\milli Ry} \\
    K2 & \SI{0.0}{\milli Ry} \\
    direction & $(0,0,1)$  \\
    \hline
    Remaining part of the material &\\
    \hline 
    K1 & \SI{-0.058}{\milli Ry} \\
    K2 & \SI{0.0}{\milli Ry} \\
    direction & $(0,0,1)$  \\
    \hline
  \end{tabular}
\\

\vspace{0.5cm}
\noindent{\bf f) Dislocation example.}\\
A D/J ratio of 20\% was chosen as a realistic value.
J and D are arbitrary, but in the order of what would be expected on a real material.
\\
\noindent

\begin{tabular}{ |l|c|  }    
    \hline
    Parameter & Value  \\
    \hline 
    Moment magnitude & \SI{2.23}{\mu_B} \\
    Temperature & \SI{0.1}{\milli\kelvin} \\
    Damping & 0.1  \\
    a & \SI{2.48}{\angstrom} \\
    $\Delta{x}/a$ & 2  \\    
    time-step & \SI{1}{\femto s} \\
    Periodicity & Y axis \\
    Crystalline structure & square \\
    \hline 
    \multicolumn{2}{|l|}{Exchange} \\
    \hline 
    Heisenberg &  \SI{1.0}{\milli Ry} \\
    DM magnitude & \SI{0.2}{\milli Ry} \\
    DM direction & \ang{90} wrt bound \\
    \hline 
    \multicolumn{2}{|l|}{Anisotropy} \\
    \hline 
    K1 & \SI{-0.04}{\milli Ry} \\
    K2 & \SI{0.0}{\milli Ry} \\
    direction & $(0,0,1)$  \\
    \hline
  \end{tabular}
\\

\section{Supplementary note S4}

\noindent{\bf Parameter transformations derivations.}\\
A procedure is described below to calculate micromagnetics solutions using the existing atomistic solver. The exchange parameters needed for an atomistic simulation to behave as the micromagnetics description are here derived step by step.

\begin{figure}[h!]
  \centering
  \def\svgwidth{3cm}
  \resizebox{0.4\textwidth}{!}{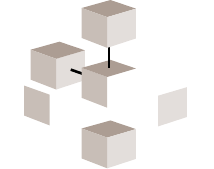}
  \caption{Finite difference stencil.}
  \label{fig:stencil}
\end{figure}

The finite-difference stencil is defined as in Fig.~\ref{fig:stencil}. The goal is to place UppASD simulation elements (considered atoms in the initial design of the tool) in the locations of finite difference nodes, and obtain from the solvers implemented in UppASD software \cite{skubic2008method} the same calculations a finite difference solver would carry out. To do so only parameters need to be tweaked. To find the relationship between parameters, let us focus on a finite difference node, $p$. The magnetic moment at the node $p$ is ${\bf  m}_p$. To make notation simpler, labels are assigned to magnetisation vectors at neighbouring space locations. The subindices $n, s, e, w, a$ and $b$ refer to positive and negative $y$, $x$ and $z$ directions, respectively. Labels are chosen after North, South, East, West, Above and Below.

\vspace{0.5cm}
\noindent{\bf a) Heisenberg exchange}\\
The Heisenberg exchange ($H_{\text{exc}}$) can be rewritten using the coordinate representation:

\begin{align}
  H_{\text{exc}} ^k &= \sum_{p=1}^{d}\sum_{q=1}^d A_{pq} \frac{\partial^2\vec m^k}{\partial x_q \partial x_p}, k=\{1,...,d\},
\end{align}
where $A_{pq}$ are the entries of the stiffness matrix $A_e$, $x_i$ is the $i^\text{th}$ coordinate, and $d$ is the number of dimensions of the space.Non-diagonal elements of $A_e$ are zeros and this leaves us with a simpler expression:

\begin{align}
  \label{eq:micro_diagonal_heis}
  H_\text{exc} ^k &= \sum_{p=1}^{d} A_{pp} \frac{\partial^2\vec m^k}{\partial x^p \partial x^p}, k={1,...,d}
\end{align}

\noindent
Using this notation, discrete approximations of second order partial derivatives of ${\bf  m}$ are of the form:
\begin{align}
  \label{eq:fd_op_xx}
  & \frac{\partial^2 \vec m}{\partial^2 x} = \frac{\vec m_e - 2\vec m_p + \vec m_w}{r_xr_x}
  \\
  \label{eq:fd_op_yy}
  & \frac{\partial^2 \vec m}{\partial y} = \frac{\vec m_n - 2\vec m_p + \vec m_s}{r_yr_y}
  \\
  \label{eq:fd_op_zz}
  & \frac{\partial^2 \vec m}{\partial z} = \frac{\vec m_a - 2\vec m_p + \vec m_b}{r_zr_z}.
\end{align}

\noindent
Thus, the Heisenberg term in the Hamiltonian must be:

\begin{equation}
\begin{aligned}
   H_\text{exc}
   & =
     \frac{A_{11}}{r_xr_x} \vec m_e 
   + \frac{A_{11}}{r_xr_x} \vec m_w
   + \frac{A_{22}}{r_yr_y} \vec m_n
   + \frac{A_{22}}{r_yr_y} \vec m_s
   + \frac{A_{33}}{r_zr_z} \vec m_a
   + \frac{A_{33}}{r_zr_z} \vec m_b
   \\
   &+ \left( \frac{-2A_{11}}{r_xr_x} + \frac{-2A_{22}}{r_yr_y} + \frac{-2A_{33}}{r_zr_z} \right) \vec m_p 
\end{aligned}
\end{equation}

\noindent
The Hamiltonian appears in a cross-product by the local magnetization in the equation of motion.
Since ${\vec m_p \times \vec m_p = \vec 0}$, the term involving the local magnetization $\vec m_p$ can be ignored.

Having written the continuum Hamiltonian exchange term in this form, it is possible to appreciate that it is equivalent to the atomistic Hamiltonian exchange when $J_{ij}$ is defined as:

\begin{align}  
  J_{ij} &= \frac{2 A_{11}}{r_xr_x} & \text{ if } i \text{and} j
  \text{ are adjacent on } x.
  \\
  J_{ij} &= \frac{2 A_{22}}{r_yr_y} & \text{ if } i \text{and} j
  \text{ are adjacent on } y.
  \\
  J_{ij} &= \frac{2 A_{33}}{r_zr_z} & \text{ if } i \text{and} j
  \text{ are adjacent on } z.
\end{align}

This means an atomistic configuration where the exchange interaction follows the vectors $J_{ij}$ here calculated would behave exactly as a micromagnetic simulation.

\vspace{0.5cm}
\noindent{\bf b)Dzyaloshinskii-Moriya}\\
To find the correspondence for Dzyaloshinskii-Moriya similar steps are taken, from the following equation:
\begin{equation}
	 \vec H^D = \nabla \cdot (D\!e \times \vec m)
\end{equation}
Thus,
\begin {equation}
  \label{eq:HDeComp}
  H_\text{De}^k= 
  \sum_{p=1}^3 \sum_{q=1}^3 \sum_{u=1}^3
  \text{De}_{pq} \xi_{kqu} \frac{\partial m^u}{\partial x^p}.
\end {equation}
\noindent
By applying the definition of the Levi-Civita symbol, the expression can be reduced to:
\begin {equation}
\begin {split}
  \label{eq:HDeCompLevCiv}
  H_\text{De}^k= 
  & \vec i \sum_{p=1}^3 \left[ \text{De}_{p2} \frac{\partial m^3}{\partial x^p}
    -\text{De}_{p3} \frac{\partial m^2}{\partial x^p} \right] +\\
  & \vec j \sum_{p=1}^3 \left[ \text{De}_{p3} \frac{\partial m^1}{\partial x^p}
    -\text{De}_{p1} \frac{\partial m^3}{\partial x^p} \right] +\\
  & \vec k \sum_{p=1}^3 \left[ \text{De}_{p1} \frac{\partial m^2}{\partial x^p}
    -\text{De}_{p2} \frac{\partial m^1}{\partial x^p} \right] 
\end {split}
\end {equation}

\pagebreak[3]
The finite difference operator is given by: 
\begin{align}
  \frac{ \partial \vec m }{\partial x^1} &= \frac{\vec m_e - \vec m_w}{2\Delta x^1}  \\
  \frac{ \partial \vec m }{\partial x^2} &= \frac{\vec m_n - \vec m_s}{2\Delta x^2}  \\
  \frac{ \partial \vec m }{\partial x^3} &= \frac{\vec m_a - \vec m_b}{2\Delta x^3} 
\end{align}

\pagebreak[2]
which can be incorporated to the following equation

\begin {equation}
\begin {split}
  \label{eq:HDeCompLevCiv1}
  H_\text{De}^k= 
  & \vec i \sum_{p=1}^3 \left[ \text{De}_{p2} \frac{\partial m^3}{\partial x^p}
    -\text{De}_{p3} \frac{\partial m^2}{\partial x^p} \right] +\\
  & \vec j \sum_{p=1}^3 \left[ \text{De}_{p3} \frac{\partial m^1}{\partial x^p}
    -\text{De}_{p1} \frac{\partial m^3}{\partial x^p} \right] +\\
  & \vec k \sum_{p=1}^3 \left[ \text{De}_{p1} \frac{\partial m^2}{\partial x^p}
    -\text{De}_{p2} \frac{\partial m^1}{\partial x^p} \right] 
\end {split}
\end {equation}
as:

\begin {equation}
  \begin {split}
    H_\text{De}^1
    &= \text{De}_{12} \frac{{m_e^3} - {m_w^3}}{2\Delta x^1}
     - \text{De}_{13} \frac{{m_e^2} - {m_w^2}}{2\Delta x^1} \\
    &+ \text{De}_{22} \frac{{m_n^3} - {m_s^3}}{2\Delta x^2}
     - \text{De}_{23} \frac{{m_n^2} - {m_s^2}}{2\Delta x^2} \\
    &+ \text{De}_{32} \frac{{m_a^3} - {m_b^3}}{2\Delta x^3}
     - \text{De}_{33} \frac{{m_a^2} - {m_b^2}}{2\Delta x^3}
  \end {split}
\end {equation} \\
\begin {equation}
  \begin {split}
    H_\text{De}^2
    &= \text{De}_{13} \frac{{m_e^1} - {m_w^1}}{2\Delta x^1}
     - \text{De}_{11} \frac{{m_e^3} - {m_w^3}}{2\Delta x^1} \\
    &+ \text{De}_{23} \frac{{m_n^1} - {m_s^1}}{2\Delta x^2}
     - \text{De}_{21} \frac{{m_n^3} - {m_s^3}}{2\Delta x^2} \\
    &+ \text{De}_{33} \frac{{m_a^1} - {m_b^1}}{2\Delta x^3}
     - \text{De}_{31} \frac{{m_a^3} - {m_b^3}}{2\Delta x^3}
  \end {split}
\end{equation} \\
\begin {equation}
  \begin {split}
    H_\text{De}^3
    &= \text{De}_{11} \frac{{m_e^2} - {m_w^2}}{2\Delta x^1}
     - \text{De}_{12} \frac{{m_e^1} - {m_w^1}}{2\Delta x^1} \\
    &+ \text{De}_{21} \frac{{m_n^2} - {m_s^2}}{2\Delta x^2}
     - \text{De}_{22} \frac{{m_n^1} - {m_s^1}}{2\Delta x^2} \\
    &+ \text{De}_{31} \frac{{m_a^2} - {m_b^2}}{2\Delta x^3}
     - \text{De}_{32} \frac{{m_a^1} - {m_b^1}}{2\Delta x^3}
  \end {split}
\end{equation}

and then expanded to:
\begin {equation}
  \begin {split}
    H_\text{De}^1 & =
    \frac{\text{De}_{12}} {2\Delta x^1} {m_e^3} -  \frac{\text{De}_{12}} {2\Delta x^1}{m_w^3} 
    - \frac{\text{De}_{13}} {2\Delta x^1} {m_e^2} +  \frac{\text{De}_{13}} {2\Delta x^1}{m_w^2}
    \\
   &+\frac{\text{De}_{22}} {2\Delta x^2} {m_n^3} -  \frac{\text{De}_{22}} {2\Delta x^2}{m_s^3} 
    - \frac{\text{De}_{23}} {2\Delta x^2} {m_n^2} +  \frac{\text{De}_{23}} {2\Delta x^2}{m_s^2}
    \\
   &+\frac{\text{De}_{32}} {2\Delta x^3} {m_a^3} -  \frac{\text{De}_{32}} {2\Delta x^3}{m_b^3} 
    - \frac{\text{De}_{33}} {2\Delta x^3} {m_a^2} +  \frac{\text{De}_{33}} {2\Delta x^3}{m_b^2}   
  \end {split}  
\end {equation}

\begin {equation}
  \begin {split}
    H_\text{De}^2 & =
    \frac{\text{De}_{13}} {2\Delta x^1} {m_e^1} -  \frac{\text{De}_{13}} {2\Delta x^1}{m_w^1} 
    - \frac{\text{De}_{11}} {2\Delta x^1} {m_e^3} +  \frac{\text{De}_{11}} {2\Delta x^1}{m_w^3}
    \\
   &+\frac{\text{De}_{23}} {2\Delta x^2} {m_n^1} -  \frac{\text{De}_{23}} {2\Delta x^2}{m_s^1} 
    - \frac{\text{De}_{21}} {2\Delta x^2} {m_n^3} +  \frac{\text{De}_{21}} {2\Delta x^2}{m_s^3}
    \\
   &+\frac{\text{De}_{33}} {2\Delta x^3} {m_a^1} -  \frac{\text{De}_{33}} {2\Delta x^3}{m_b^1} 
    - \frac{\text{De}_{31}} {2\Delta x^3} {m_a^3} +  \frac{\text{De}_{31}} {2\Delta x^3}{m_b^3}   
  \end {split}  
\end {equation}

\begin {equation}
  \begin {split}
    H_\text{De}^3 & =
    \frac{\text{De}_{11}} {2\Delta x^1} {m_e^2} -  \frac{\text{De}_{11}} {2\Delta x^1}{m_w^2} 
    - \frac{\text{De}_{12}} {2\Delta x^1} {m_e^1} +  \frac{\text{De}_{12}} {2\Delta x^1}{m_w^1}
    \\
   &+\frac{\text{De}_{21}} {2\Delta x^2} {m_n^2} -  \frac{\text{De}_{21}} {2\Delta x^2}{m_s^2} 
    - \frac{\text{De}_{22}} {2\Delta x^2} {m_n^1} +  \frac{\text{De}_{22}} {2\Delta x^2}{m_s^1}
    \\
   &+\frac{\text{De}_{31}} {2\Delta x^3} {m_a^2} -  \frac{\text{De}_{31}} {2\Delta x^3}{m_b^2} 
    - \frac{\text{De}_{32}} {2\Delta x^3} {m_a^1} +  \frac{\text{De}_{32}} {2\Delta x^3}{m_b^1}   
  \end {split}  
\end {equation}

\noindent
The atomistic expression for the DM term is:
\begin {equation}
  H_{\text{D}_i} = \sum_j{\vec D_{ij}}\times {\vec s_j}
\end {equation}

\noindent
Consider here that atom $j$ is instead the finite difference node $m_p$.
If we also consider the neighbours in the finite difference as neigbours of the atom, we would get:

\begin {equation}
  \begin{split}
    H_D^1
    & = {D_w^2}{m_w^3} - {D_w^3}{m_w^2} 
    + {D_e^2}{m_e^3} - {D_e^3}{m_e^2} \\
    & + {D_s^2}{m_s^3} - {D_s^3}{m_s^2} 
    + {D_n^2}{m_n^3} - {D_n^3}{m_n^2} \\
    & + {D_b^2}{m_b^3} - {D_b^3}{m_b^2} 
    + {D_am^2}{m_a^3} - {D_a^3}{m_a^2}  
  \end{split}
\end{equation}
\begin{equation}
 \begin{split}
    H_D^2
    & = {D_w^3}{m_w^1} - {D_w^1}{m_w^3} 
    + {D_e^3}{m_e^1} - {D_e^1}{m_e^3} \\
    & + {D_s^3}{m_s^1} - {D_s^1}{m_s^3} 
    + {D_n^3}{m_n^1} - {D_n^3}{m_n^3} \\
    & + {D_b^3}{m_b^1} - {D_b^1}{m_b^3} 
    + {D_a^3}{m_a^1} - {D_a^1}{m_a^3}  
  \end{split}
\end {equation}
\begin{equation}
 \begin{split}
    H_D^3
    & = {D_w^1}{m_w^2} - {D_w^2}{m_w^1} 
    + {D_e^1}{m_e^2} - {D_e^2}{m_e^1} \\
    & + {D_s^1}{m_s^2} - {D_s^2}{m_s^1} 
    + {D_n^1}{m_n^2} - {D_n^2}{m_n^1} \\
    & + {D_b^1}{m_b^2} - {D_b^2}{m_b^1} 
    + {D_a^1}{m_a^2} - {D_a^2}{m_a^1}  
  \end{split}
\end {equation}

\noindent
Then, a correspondence between $\text{De}_{ij}$ and $D_*^j$ can be found:
\begin{equation}
 \begin{aligned}
   D_w^1 = \frac{-D\!e_{11}}{2\Delta x^1} && 
   D_w^2 = \frac{-D\!e_{12}}{2\Delta x^1} &&
   D_w^3 = \frac{-D\!e_{13}}{2\Delta x^1} &&
   \\
   D_e^1 = \frac{ D\!e_{11}}{2\Delta x^1} && 
   D_e^2 = \frac{ D\!e_{12}}{2\Delta x^1} &&
   D_e^3 = \frac{ D\!e_{13}}{2\Delta x^1} &&
   \\
   D_s^1 = \frac{-D\!e_{21}}{2\Delta x^2} &&
   D_s^2 = \frac{-D\!e_{22}}{2\Delta x^2} &&
   D_s^3 = \frac{-D\!e_{23}}{2\Delta x^2} &&
   \\
   D_n^1 = \frac{ D\!e_{21}}{2\Delta x^2} &&
   D_n^2 = \frac{ D\!e_{22}}{2\Delta x^2} &&
   D_n^3 = \frac{ D\!e_{23}}{2\Delta x^2} &&
   \\
   D_b^1 = \frac{-D\!e_{31}}{2\Delta x^3} &&
   D_b^2 = \frac{-D\!e_{32}}{2\Delta x^3} &&
   D_b^3 = \frac{-D\!e_{33}}{2\Delta x^3} &&
   \\
   D_a^1 = \frac{ D\!e_{31}}{2\Delta x^3} &&
   D_a^2 = \frac{ D\!e_{32}}{2\Delta x^3} &&
   D_a^3 = \frac{ D\!e_{33}}{2\Delta x^3} &&
   \\
 \end{aligned}
\end{equation}

\section{Supplementary note S5}
 A pulsed field is used to generate locally a strong magnetic field of 600 T in the $-z$ direction. This causes a fliping of spins of atoms in a small region of the simulation box. Such strong fields can be achieved by e.g. a Laser via the inverse Faraday effect \cite{kimel}, and enable the creaton of skyrmions. The pulsed field used to create the skyrmions, has a radius of 10 lattice spacing (\SI{27.15}{\angstrom} in total) and lasted \SI{2}{\pico s}. After the pulse was applied, moments in the regions affected quickly flip their spin. Neighbouring magnetic moments rotate to reduce the high exchange energy introduced by the abrupt change in magnetization direction, and a skrmion state is stabilized.

\section{Supplementary note S6}

Atoms tend to keep a constant distance from each other in the lattice. As an atom is displaced, its neighbours pull it back, trying to preserve the inter-atomic distance. The resulting behaviour is similar to that of a mesh of springs or elastic bands. An added constraint is that the atoms surrounding the dislocation must minimize the potential energy with respect to the rest of neighbouring atoms.  This is a requirement since the continuum cannot adapt to the distorted atomistic lattice, and the transition atoms need to behave in the same way as the continuum. Placing the atoms is then formulated as an optimization problem, where the atoms at the boundaries of the domain are locked to lattice points and one row of atoms is removed in the interior. The target function is the elastic potential energy for the configuration of atoms. Assuming the material is isotropic, the potential energy can be written as

  \begin{equation}
    E = \sum_{i} \frac12 k (\Delta r_i)^2,
  \end{equation}
  where $k$ is a material parameter and $\Delta p_i$ is the displacement of the atom from its rest position.
  The positions are found using an iterative gradient descent algorithm on the energy while preserving the constrained atoms in place. Most of the atomic distances are very close to the lattice parameter. Around 97.5\% of the bonds are shorter than $1.09a$, but the longest bond is of $1.36a$. As some atoms were removed, no bond between atoms is shorter than $1a$. Varying bond distances have an effect in the interaction strengths. For Heisenberg exchange, a polynomial is fitted to the values of a known material with cubic (bcc) lattice. The material used as reference is iron-bcc. The calculation of Dzyaloshinsky-Moriya interaction vectors is a bit more cumbersome. The Levy-Fert mechanism can be used to model the DM for some materials, but the topology of the surrounding layers is needed \cite{PhysRevLett.44.1538}. For the sake of demonstration, a much more simplified model was used. The DM vectors chosen are coplanar to the material, parallel to the bond and their magnitude is proportional to the distortion of the length. The magnitude of the DM vectors when the link is not distorted was chosen to be 20\% of the Heisenberg exchange for first neighbours. A script was written in order to generate this configuration. The values are calculated for $J_{ij}$ and $\vec D_{ij}$ as
  \begin{align}  
    J_{ij} = &J_0 (-0.05988\|r_{ij}\|^3 + 0.95532 \|r_{ij}\|^2 \nonumber \\
    &-4.9516 \|r_{ij}\|  + 8.31778) \\
    \vec D_{ij} =& 0.2 J_0 \frac{R \vec r_{ij}}{\|r_{ij}\|} \left(2 - \frac{\|r_{ij}\|}{a}\right)
  \end{align}
  where $J_0 = \SI{1}{\milli Ry}$ is the scale of the exchange, and R is either the identity matrix when the DM vectors lying along the bond connecting atoms, or a 90 degree rotation matrix for DM vectors lying perpendicular to  the bond. The evaluation of the polynomial for $J_{ij}$ gives a value of 1 when $\|r_{ij}\|$ is exactly one lattice spacing (\SI{2.48}{\angstrom}).